\documentclass[fleqn,usenatbib]{mnras}

\usepackage{newtxtext,newtxmath}
\usepackage[T1]{fontenc}

\DeclareRobustCommand{\VAN}[3]{#2}
\let\VANthebibliography\thebibliography
\def\thebibliography{\DeclareRobustCommand{\VAN}[3]{##3}\VANthebibliography}

\usepackage{subcaption}
\usepackage{longtable}
\usepackage{tablefootnote}
\usepackage{threeparttable}
\usepackage{multirow}
\usepackage{graphicx}	% Including figure files
\usepackage{amsmath}	% Advanced maths commands
\usepackage[table]{xcolor}
\usepackage{cleveref}
\usepackage{xcolor}
\newcommand{\hd}{HD\,143006\,}
\title[\hd]{HD 143006: Interferometric Confirmation of Misaligned Protoplanetary Disc with CHARA/MIRCX and VLTI/PIONIER}

\author[I. Codron et al.]{
I. Codron,$^{1}$\thanks{E-mail: ic302@exeter.ac.uk (IC)}
S. Kraus,$^{1}$
J. D. Monnier,$^{2}$
S. Marino,$^{1}$
C. L. Davies,$^{1}$
N. Anugu,$^{3}$
T. Gardner,$^{1}$
\newauthor
N. Ibrahim,$^{2}$
C. Lanthermann,$^{3}$
J-B Le Bouquin$^{4}$
\\
$^{1}$University of Exeter, School of Physics and Astronomy, Astrophysics Group, Stocker Road, Exeter, EX4 4QL, UK\\
$^{2}$Astronomy Department, University of Michigan, Ann Arbor, MI 48109, USA\\
$^{3}$The CHARA Array of Georgia State University, Mount Wilson Observatory, Mount Wilson, CA 91203, USA\\
$^{4}$Institut de Planetologie et d'Astrophysique de Grenoble, Grenoble 38058, France\\
}

\date{Accepted 2025/06/24}

\pubyear{2025}

\begin{document}
\label{firstpage}
\pagerange{\pageref{firstpage}--\pageref{lastpage}}
\maketitle

% Abstract of the paper
\begin{abstract}
The outer regions of the protoplanetary disc surrounding the T Tauri star \hd show rings, dust asymmetries and shadows. Whilst rings and dust asymmetries can arise from companions and other mechanisms, shadows and misaligned discs in particular are typically attributed to the presence of misaligned planets or stellar-mass companions. To understand the mechanisms that drive these traits, the innermost regions of discs need to be studied. Using CHARA/MIRCX and VLTI/PIONIER, we observed the sub-au region of \hd. We constrain the orientation of the inner disc of \hd and probe whether a misalignment between the inner and outer disc could be the cause of the shadows. Modelling
the visibilities using a geometric model, the inclination and position angle are found to be $i=22^\circ\pm 3^\circ$ and $\mathrm{PA}=158^\circ\pm 8^\circ$ respectively, with an inner dust sublimation radius of $\sim0.04$ au. The inner disc is misaligned by $39^\circ\pm4^\circ$ with respect to the outer disc, with the far side of the inner disc to the east and the far side of the outer disc to the west. We constrain $h/R$ (scattering surface/radius of scattered light) of the outer disc at $18$ au to be about $13\%$ by calculating the offset between the shadow position and the central star. No companion was detected, with a magnitude contrast of $4.4$ in the H-band and placing an upper mass limit of $0.17 M_\odot$ at separations of $0-8$ au. Therefore, we cannot confirm or rule out that a low-mass star or giant planet is responsible for the misalignment and dust sub-structures.

\end{abstract}

% Select between one and six entries from the list of approved keywords.
% Don't make up new ones.
\begin{keywords}
stars:pre-main-sequence -- protoplanetary discs -- object:\hd
\end{keywords}

%%%%%%%%%%%%%%%%%%%%%%%%%%%%%%%%%%%%%%%%%%%%%%%%%%

%%%%%%%%%%%%%%%%% BODY OF PAPER %%%%%%%%%%%%%%%%%%

\section{Introduction}
\label{sec:intro}
Over the last decade or so, we have seen the number of exoplanet discoveries increase greatly (eg. \citealt{shweta2024MNRAS.531.4464D}), as well as a large diversity in these discoveries (eg. \citealt{jontof2019AREPS..47..141J,eberlein2024arXiv240720117E,levi2024arXiv240800070W}). The process of planet formation is still far from being completely understood, and the very inner region of these protoplanetary discs may provide us with information on the formation and migration of the ubiquitous population of close-in planets \citep{2023ASPC..534..539M}.

Using the Very Large Telescope Interferometer (VLTI) and the Center for High Angular Resolution Astronomy (CHARA) Array, these inner regions can be spatially resolved and dynamical processes that are happening at the inner edge of the disc can be observed. 

Whilst the outer disc regions emit strongly in the far-infrared and sub-mm wavelengths due to their low temperature, the inner disc region is bright in the near-infrared (NIR) and mid-infrared wavelengths. The very inner disc region ($0.1$-$1$ au) is where the sublimation rim lies - a region where the temperature exceeds that at which dust can survive (e.g. $\sim1500$K, \citealt{monniermillan2002, dullemond2010ARA&A..48..205D}) and the disc becomes purely gaseous.
% By investigating these inner regions and comparing the size scales of the innermost rim, we can infer the evolutionary stage that the disc is in. 

As protoplanetary discs evolve over time, two particularly interesting points in their evolution are the `pre-transitional' and `transitional' disc stages \citep{benisty2018A&A...619A.171B,transition2014prpl.conf..497E}. First identified by their spectral energy distributions (SEDs), these discs show a small amount, or lack of excess in the near-infrared (NIR), but significant excess is seen in the mid- and far-infrared. This dip in the SEDs indicates that there is an inner dust cavity. Numerous techniques can be used to probe these cavities; such as (sub)millimeter interferometric imaging \citep{dong2017ApJ...836..201D,vdm2021AJ....161...33V}, mid-infrared interferometry \citep{kraus2013ApJ...768...80K,kluska2018ApJ...855...44K}, and scattered light images \citep{aven2018ApJ...863...44A,benisty2018A&A...619A.171B,bohn2022A&A...658A.183B}. Some (pre-)transitional discs present asymmetries seen in scattered-light and thermal light which can potentially be explained by a misaligned inner disc - seen in the form of narrow shadow lines in the outer disc \citep{2015marino,benisty2017A&A...597A..42B,pinilla2015A&A...584L...4P}, or low-amplitude azimuthal variations \citep{debes2017ApJ...835..205D}. Additional evidence of misalignments and cavities can be found in the form of deviations from Keplerian rotation seen in gas lines \citep{Loomis2017,Boehler2018} and ``dipper" stars \citep{bodman2017MNRAS.470..202B}.
% Dipper stars are a fairly common subclass of young stars ($\sim20\%-30\%$) \citep{alencar2010A&A...519A..88A,cody2014AJ....147...82C} that exhibit photometric variability during short duration extinction events. It has been proposed that these photometric dips might be caused by transiting circumstellar material clumps related to planetesimal formation, which would require a highly inclined orbit \citep{bodman2017MNRAS.470..202B}. 
% However, sub-mm observations have shown that there is a wide range of inclinations, including face-on \citep{ansdell2016MNRAS.462L.101A, ansdell2018ApJ...859...21A}, providing evidence that disc misalignments may be common in young stars. Periodic dipper stars are thought to cause a warp due to a strong dipolar stellar magnetic field being misaligned with respect to the disc midplane, inducing a tilt in the inner disc \citep{bouvier2007A&A...463.1017B}. 
Disc warping is thought to be caused by the gravitational interaction from massive companions. This theory lends itself well to transition discs due to the SED implying a dust-depleted cavity which can be carved out by multiple planets \citep{bae2019ApJ...884L..41B} or stellar companions \citep{price2018MNRAS.477.1270P,aly2020MNRAS.492.3306A}. 

% Whilst at times the term transitional is used interchangeably between transitional and pre-transitional discs, the difference pertinent to this work, lies in the location of the inner rim of the cavity. Particularly, \cite{espaillat2010ApJ...717..441E} found that when modelling the inner region of pre-transitional discs, the innermost wall is located at the dust-destruction radius. Both SEDs and modelling can therefore allow for a more accurate determination of these objects in their evolutionary timescales. ugh this sounds so bleh

This work focuses on the disc surrounding \hd. 
\hd is a T Tauri  G$7$ star, with an effective temperature  $\mathrm{T}_{\mathrm{eff}}=5880$K, bolometric luminosity $L_*=4.58L_\odot$, and stellar mass $M_*=1.8\substack{+0.2 \\ -0.3}M_\odot$ \citep{andrews2018ApJ...869L..41A}. \hd is located at a distance of $165\pm5$ pc in the Upper Sco region \citep{gaia2018A&A...616A...1G} with an age of $5$-$11$\,Myr \citep{pecaut2012ApJ...746..154P}. ALMA observations carried out by \cite{barenfeld2016ApJ...827..142B} resolved the outer disc and showed a centrally depleted cavity, and the ALMA DSHARP survey \citep{andrews2018ApJ...869L..41A,perez2018ApJ...869L..50P} carried out ALMA high-resolution observations ($0.05''$), showing structural features such as rings and gaps. Polarised scattered light images in the J-band from VLT/SPHERE observations, with an angular resolution of $\sim0.037''$, showed the presence of rings and gaps in the disc of \hd as well as a large-scale asymmetry \citep{benisty2018A&A...619A.171B}.
Three bright concentric rings have been detected in the dust continuum emission \citep{perez2018ApJ...869L..50P} with the innermost ring detected with ALMA having a radius of 8\, au ($0.048$"), and the other two rings at $40$\, au and $64$\, au. \citet{perez2018ApJ...869L..50P} find an inclination ($i$) and position angle (PA) of $24.1^\circ\pm1.0^\circ$ and $164.3^\circ\pm2.4^\circ$ respectively for the ring at 8\, au. For the outer disc at $64$ au they find $i=17.02^\circ\pm0.14^\circ$ and $\mathrm{PA}=176.2^\circ\pm0.6^\circ$. \citet{benisty2018A&A...619A.171B} use the ALMA moment 1 map of the $^{12}$CO line to obtain a very similar outer disc inclination of $17.5^\circ\pm0.2^\circ$ and PA of $169.8^\circ\pm0.5^\circ$. The NIR excess for {\hd} indicates that hot dust exists at the sublimation radius of $\sim0.1$ au, whereby \citet{laz2017A&A...599A..85L} fit a geometrical model with $i=27^\circ\pm4^\circ$ and $\mathrm{PA}=180^\circ\pm13^\circ$. It is thought that a misalignment between the innermost ring ($<1$ au) and the outer disc is causing the shadow seen in scattered light. However, it has been impossible to resolve this emission in scattered light or thermal emission due to the resolution of direct imaging and sub-mm interferometry.
The origin of the observed structures of this object is not yet understood. 
Whilst multiple features have been identified (see \citealt{benisty2018A&A...619A.171B,perez2018ApJ...869L..50P}), a particularly interesting feature are the narrow shadows along the north-south direction, discovered in IR scattered light by \citet{benisty2018A&A...619A.171B}. These shadows form the focal point of this work.
%however, it has not been possible to resolve structures closer-in in scattered light or thermal emission at optical to mm wavelengths.

A recent study by \cite{Ballabio2021MNRAS.504..888B} tried to reproduce the disc features seen in scattered light by performing 3D hydrodynamic simulations in which they considered three different scenarios: i) a stellar companion inclined at $60^\circ$ to the disc plane, ii) a misaligned planet around a single star, and iii) an inclined stellar companion with a planetary companion. They concluded that a circumbinary inclined stellar companion (mass ratio $=0.2$, separation $2$ au) with an embedded planet ($10 M_J$ at $32$ au) was the only scenario that successfully reproduced the main features seen inside $40$\, au in the scattered light image, channel maps, and ALMA observations. 
%$166\substack{+0.41 \\ -0.50}$ pc in the Upper Sco region \citep{gaia2023A&A...674A...1G, gaia2016A&A...595A...1G,gaia2023A&A...674A..32B}
%\cite{laz2017A&A...599A..85L} also find a position angle and inclination angle of X and Y respectively. 

% Investigating the origin of the misalignments in \hd further requires resolving the inner disc regions down to the sublimation radius. This work uses the unique high-angular resolution imaging capabilities of the CHARA Array to resolve the inner au of \hd, determine if there is a misalignment between the inner and outer disc, and to search for a stellar companion located inside the cavity.
Herein, we characterise the NIR circumstellar emission using CHARA/MIRCX and VLTI/PIONIER data. Constraining the innermost disc geometry and orientation can help us to measure and confirm if the inner disc is misaligned relative to the outer disc. Finally, building on the simulations by \cite{Ballabio2021MNRAS.504..888B}, we investigate the hypothesis of a stellar companion in a circumbinary disc.  
This paper is structured as follows. In Sect. \ref{sec:obs} we present our observations and data reduction, in Sect. \ref{sec:models} we present our geometrical modelling and disc misalignment calculations, in Sect. \ref{sec:results} we present and discuss our findings and summarise our results in Sect. \ref{sec:conclusions}.

\section{Observations and Data Reduction}
\label{sec:obs}

The near-infrared interferometric data used in this work were obtained using the CHARA Array and the VLTI. The data spans a period of eight years, combining new observations with archival data. Information about the observations can be found in Table \ref{tab:hd143006obs}. A typical observation comprises of a calibrator-science-calibrator sequence or `block', with some nights having multiple sequential blocks. The raw visibilities are calibrated using calibrator stars selected using SearchCal\footnote{\url{https://www.jmmc.fr/english/tools/proposal-preparation/search-cal/}} \citep{chelli2016A&A...589A.112C}.

% \begin{table*}
%     \caption{\label{tab:hd143006obs} Full list of interferometric observations of \hd. Data from programme 190.C-0963 are lacking calibrator information, due to being taken pre-calibrated from the JMMC OiDB.}
%     \centering
%     \begin{tabular}{l|c|r|c|c|c}
%     \hline
%     Date & Programme ID & Array Config & Instrument & $\Delta\lambda/\lambda$ & Calibrator(s) used\\
%     \hline
%     2013-06-06 & 190.C-0963(D) & A1-G1-K0-J3 & PIONIER &  15 & - \\
%     2013-06-17 & 190.C-0963(E) & D0-H0-G1-I1 & PIONIER &  15 & - \\
%     2013-07-04 & 190.C-0963(F) & A1-B2-C1-D0 & PIONIER &  15 & - \\
%     \hline
%     2019-07-20 & 0103.C-0915(A) & A0-G1-J2-J3 & PIONIER &   \&  & HD\,146235  \\
%     \multicolumn{1}{c|}{"}      & 098.C-0910(A) & A0-G1-J2-J3 & GRAVITY &  22 \& 500 & HD\,154436 \\

%     2017-04-27\,\tablefootmark{a} & 098.C-0910(A) & A0-G1-J2-J3 & GRAVITY & 22 \& 4000 & HD\,57087  \\
%     2017-04-28\,\tablefootmark{a} & 098.C-0910(A) & A0-G1-J2-J3 & GRAVITY & 22 \& 500 & HD\,49647 \\
%     2018-01-11 & 098.C-0910(A) & A0-G1-J2-J3 & GRAVITY &  22 \& 4000 & HD\,49647 \\
%     2018-02-06 & 098.C-0910(A) & A0-G1-J2-J3 & GRAVITY &  22 \& 4000 & HD\,38117 \\ 
%     \multicolumn{1}{c|}{"}      & 098.C-0910(A) & A0-G1-J2-J3 & GRAVITY &  22 \& 500 & HD\,55137 \\
%     \hline

%     \hline
%     \end{tabular}
%     \tablefoot{
%     \tablefoottext{a}{Observations on consecutive days were grouped into one epoch for continuum analysis.} \\
%     }
% \end{table*}

\begin{table*}
    \caption{\label{tab:hd143006obs} Full list of interferometric observations of \hd with calibrator stars taken from SearchCal \citep{chelli2016A&A...589A.112C}. Data from programme 190.C-0963 are lacking calibrator information, due to being taken post-calibrated from the JMMC OiDB. UD refers to the uniform-disc diameter.}
    \centering
    \begin{threeparttable}
        % \begin{tabular}{l|c|r|c|c|c}
        % \hline
        % Date & Programme ID & Array Config & Instrument & Calibrator(s) used & UD (mas)\\
        % \hline
        % 2013-06-06 & 190.C-0963(D) & A1-G1-K0-J3 & PIONIER &  - & -\\
        % 2013-06-17 & 190.C-0963(E) & D0-H0-G1-I1 & PIONIER &  - & -\\
        % 2013-07-04 & 190.C-0963(F) & A1-B2-C1-D0 & PIONIER &  - & -\\
        % 2019-07-20 & 0103.C-0915(A) & A0-G1-J2-J3 & PIONIER   & HD\,146235  & $0.186\pm0.005$ \\
        % \multicolumn{1}{c|}{"}      &  &  &  &  HD\,154436 & $0.489\pm0.003$\\
        % 2020-06-22& 2020A-M5 & X-W2-W1-S2-S1-E2 & MIRC-X & HD\,144114  & $0.322\pm0.008$\\
        % \multicolumn{1}{c|}{"}      &  &  &  &  HD\,139575 & $0.398\pm0.009$\\
        % 2021-05-12& 2021A-M7 & E1-W2-W1-S2-X-E2 & MIRC-X & HD\,144114 &$0.322\pm0.008$\\
        % \multicolumn{1}{c|}{"}      &  &  &  &  HD\,139575 &$0.398\pm0.009$\\
        % 2021-05-13& 2021A-M7 & X-W2-W1-S2-S1-E2 & MIRC-X & HD\,151259 &$0.418\pm0.011$\\
        % \multicolumn{1}{c|}{"}      &  &  &  &  HD\,145965 &$0.209\pm0.005$\\
        % \multicolumn{1}{c|}{"}      &  &  &  &  HD\,139487 &$0.305\pm0.007$\\
        % 2021-05-14& 2021A-M7 & E1-W2-W1-S2-X-E2 & MIRC-X & HD\,135969 &$0.211\pm0.006$\\
        % 2021-08-06& 105.20T6.002 & K0-G2-D0-J3 & PIONIER & HD\,143132  & $0.199\pm0.005$\\
        % 2021-08-07& 105.20T6.002 & K0-G2-D0-J3 & PIONIER &  HD\,143132 &$0.199\pm0.005$\\
        % \multicolumn{1}{c|}{"}      &  &  &  &  HD\,143647 &$0.245\pm0.006$\\
        % \multicolumn{1}{c|}{"}      &  &  &  &  HD\,143049 & $0.251\pm0.006$\\
        % 2021-08-08& 105.20T6.002 & K0-G2-D0-J3 & PIONIER & HD\,143647  &$0.245\pm0.006$\\
        % \multicolumn{1}{c|}{"}      &  &  &  &  HD\,317458 & $0.289\pm0.008$\\
        % \hline
        % \hline
        % \end{tabular}
        \begin{tabular}{l|c|c|r|c|c}
        \hline
        Date & Programme ID & Instrument & Array Config & Calibrator(s) used & UD (mas)\\
        \hline
        2013-06-06 & 190.C-0963(D) & PIONIER & A1-G1-K0-J3 & - & -\\
        2013-06-17 & 190.C-0963(E) & PIONIER & D0-H0-G1-I1 & - & -\\
        2013-07-04 & 190.C-0963(F) & PIONIER & A1-B2-C1-D0 & - & -\\
        2019-07-20 & 0103.C-0915(A) & PIONIER & A0-G1-J2-J3 & HD\,146235  & $0.186\pm0.005$ \\
        \multicolumn{1}{c|}{"}      &  &  &  &  HD\,154436 & $0.489\pm0.003$\\
        2020-06-22 & 2020A-M5 & MIRC-X & X-W2-W1-S2-S1-E2 & HD\,144114  & $0.322\pm0.008$\\
        \multicolumn{1}{c|}{"}      &  &  &  &  HD\,139575 & $0.398\pm0.009$\\
        2021-05-12 & 2021A-M7 & MIRC-X & E1-W2-W1-S2-X-E2 & HD\,144114 &$0.322\pm0.008$\\
        \multicolumn{1}{c|}{"}      &  &  &  &  HD\,139575 &$0.398\pm0.009$\\
        2021-05-13 & 2021A-M7 & MIRC-X & X-W2-W1-S2-S1-E2 & HD\,151259 &$0.418\pm0.011$\\
        \multicolumn{1}{c|}{"}      &  &  &  &  HD\,145965 &$0.209\pm0.005$\\
        \multicolumn{1}{c|}{"}      &  &  &  &  HD\,139487 &$0.305\pm0.007$\\
        2021-05-14 & 2021A-M7 & MIRC-X & E1-W2-W1-S2-X-E2 & HD\,135969 &$0.211\pm0.006$\\
        2021-08-06 & 105.20T6.002 & PIONIER & K0-G2-D0-J3 & HD\,143132  & $0.199\pm0.005$\\
        2021-08-07 & 105.20T6.002 & PIONIER & K0-G2-D0-J3 & HD\,143132 &$0.199\pm0.005$\\
        \multicolumn{1}{c|}{"}      &  &  &  &  HD\,143647 &$0.245\pm0.006$\\
        \multicolumn{1}{c|}{"}      &  &  &  &  HD\,143049 & $0.251\pm0.006$\\
        2021-08-08 & 105.20T6.002 & PIONIER & K0-G2-D0-J3 & HD\,143647  &$0.245\pm0.006$\\
        \multicolumn{1}{c|}{"}      &  &  &  &  HD\,317458 & $0.289\pm0.008$\\
        \hline
    \end{tabular}

        \footnotesize{The `X' in array configuration signifies a missing telescope}        
    \end{threeparttable}
\end{table*}

\subsection{CHARA}

The CHARA Array located on Mt. Wilson, California, is a Y-shaped array of six $1$-meter telescopes, with a maximum baseline length (B) of $331$\,metres. The instrument used to take observational data is the Michigan Infrared Combiner - eXeter (MIRC-X), which is a highly sensitive 6-telescope beam combiner, lending a maximum angular resolution of $\lambda/2B = 0.51$mas in the H-band ($\lambda=1.65\mu$m) \citep{2018SPIE10701E..23K,2020AJ....160..158A}.
The MIRC-X instrument was used in the H-band mode with a spectral resolution of $\mathcal{R}=50$. Observations were carried out using five out of the six telescopes (5T) due to low elevation in the $2020$A-M$5$ and $2021$A-M$7$ programmes (see Table \ref{tab:hd143006obs}). The data were reduced using the MIRC-X pipeline version $1.3.5$\footnote{Available at \url{https://gitlab.chara.gsu.edu/lebouquj/mircx_pipeline}}, and calibrated using a custom IDL routine\footnote{Contact monnier@umich.edu for IDL routine}. The pipeline produces science-ready visibilities and closure phases in OIFITS format \citep{anugu2020SPIE11446E..0NA,pauls2005PASP..117.1255P}. 

\subsection{VLTI}
Data was recorded as part of the observing programmes 0103.C-0915(A) and 105.20T6.001. We employed PIONIER in the H-band with $6$ spectral channels and a resolution of $\mathcal{R}=40$. The data was reduced with the reduction pipeline \texttt{pndrs v3.52} \citep{pndrs2011A&A...535A..67L}. Programmes 0103.C-0915(A) and 105.20T6.001 used the $1.8$\,metre Auxiliary Telescopes (AT) in the large ($B\sim120$m) and medium ($B\sim100$m) configuration respectively.
% 105 was using AT telescopes medium config, 0103 also AT but large config (120m). 105 is 100m baseline
We also included archival data from programme 190.C-0963 \footnote{Taken from \url{http://oidb.jmmc.fr/index.html}}, which has already been published as part of the survey by \citet{laz2017A&A...599A..85L}. These data featured 3 spectral channels with a resolution of $\mathcal{R}=15$. 

\subsection{Reduced Data}

% \begin{figure}
%     \centering
%     \includegraphics[width=\linewidth]{Images/uv.pdf}
%     \caption{Caption}
%     \label{fig:enter-label}
% \end{figure}

% \begin{figure*}
% \centering
% \begin{subfigure}[b]{.45\linewidth}
% \includegraphics[width=\linewidth]{Images/uv.pdf}
% %\caption{A gull}\label{fig:gull}
% \end{subfigure}

% \centering
% \begin{subfigure}[b]{\linewidth}
% \includegraphics[width=\linewidth]{Images/FEBraw.pdf}
% %\caption{A mouse}\label{fig:mouse}
% \end{subfigure}

% \caption{\textit{Top: }(u,v) coverage of the combined data for all epochs on both instruments. \textit{Bottom: } Visibility and closure phase (T3PHI) of the combined data at all epochs with both PIONIER and MIRC-X instruments. No post-calibration data cleaning has been applied to the visibility or closure phase data in these plots.}
% \label{fig:rawdata}
% \end{figure*}

\begin{figure}
\centering
\begin{subfigure}[b]{\linewidth}
\includegraphics[width=0.85\linewidth]{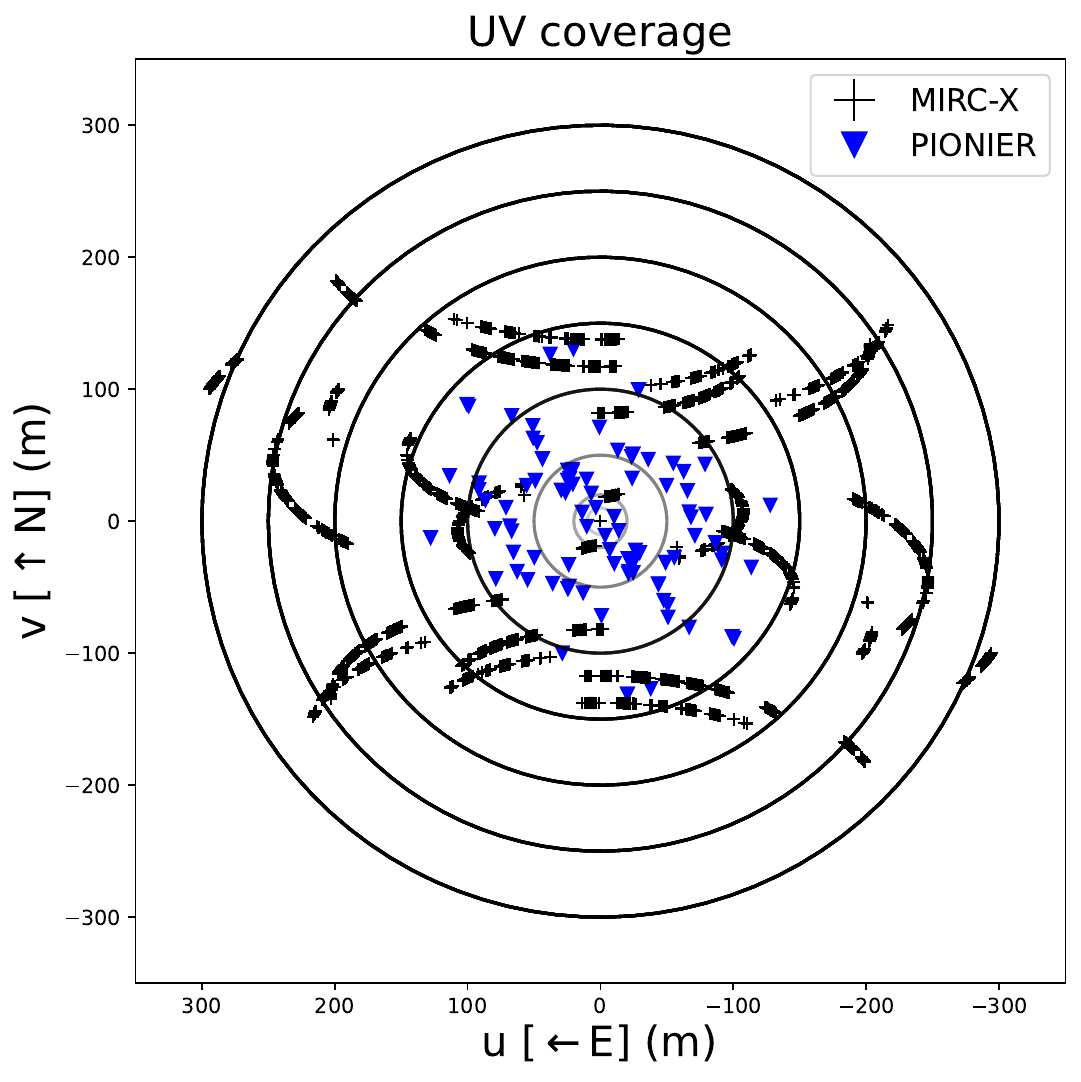}
%\caption{A gull}\label{fig:gull}
\end{subfigure}

\centering
\begin{subfigure}[b]{\linewidth}
\includegraphics[width=\linewidth]{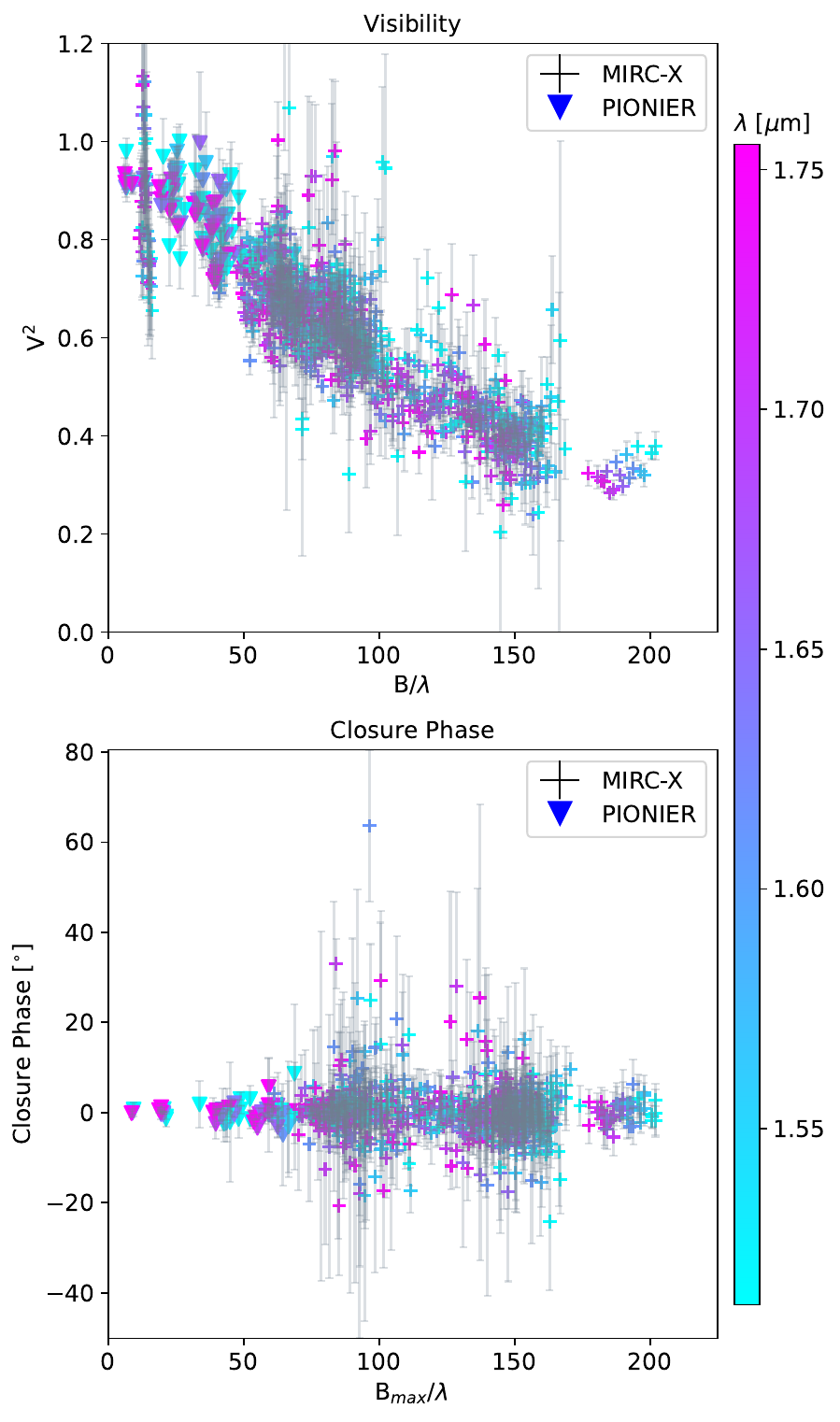}
%\caption{A mouse}\label{fig:mouse}
\end{subfigure}

\caption{\textit{Top: }(u,v) coverage of the combined data for all epochs on both instruments. \textit{Middle: } Visibility of the combined data at all epochs with both PIONIER and MIRC-X instruments. \textit{Bottom: }Closure phase (T3PHI) of the combined data at all epochs with both PIONIER and MIRC-X instruments. No post-calibration data cleaning has been applied to the visibility or closure phase data in these plots.}
\label{fig:rawdata}
\end{figure}

\subsubsection{Visibilities}
From an initial view of the combined instrument visibility data in the middle panel of Fig \ref{fig:rawdata}, there are characteristics and features that are expected. The curve has a gentle downward slope rather than a steep drop, indicating that the disc is small. Compact emission in the H-band has already been observed by \cite{laz2017A&A...599A..85L} but only marginally resolved. The data points do not show a significant vertical spread, implying the disc is not highly inclined. At short baselines, there appear to be a large number of data points falling below a visibility of $1$, which may signify that there is a large over-resolved halo.

We see a chromatic effect in our visibility data as evidenced in Figure \ref{fig:rawdata}. We observe a wavelength dependence on the observed visibilities that becomes more prominent towards longer wavelengths, as evidenced by the upward `arc' of data points (at higher spatial frequencies). This comes from us probing the H-band region where the SED transitions from the star-dominated to the disc-dominated region. At shorter wavelengths, the star is brighter and is therefore contributing more to the SED, whilst at longer wavelengths the disc is brighter. We see the chromatic effect much more at longer baselines as this equates to a lower spatial frequency and lower visibilities \citep{Kluska2014A&A...564A..80K}. 

\subsubsection{Closure Phases}
The closure phase (CP) measurements from both instruments are seen in the bottom panel of Fig \ref{fig:rawdata}. The majority of data points appear to be $<10^\circ$, with a high density of data points $\sim0^\circ$. This suggests a symmetrical object and there is no immediate indication of a companion.

\subsection{Data Cleaning}
After data reduction and data calibration, we also set upper limits on error bars, removing points that exceeded these limits. For the visibility data, we employ an error bar cut-off at $\mathcal{V}_\mathrm{err}=0.2$, and closure phase error cut-off at $\mathrm{CP_{err}}=10^\circ$. The features discussed above become more prominent, in Figures \ref{fig:ringachrom} and \ref{fig:ringchromaticVISCP}.

\section{Modelling}
\label{sec:models}

\subsection{Inner Disc Geometry}

The interferometric visibility and closure phase data points were fitted using the geometric modelling code PMOIRED (Parametric Modeling of Optical InteRferomEtric Data,  \citealt{pmoired2022SPIE12183E..1NM}). We estimate the stellar radius of \hd to be $0.06$ mas, which is significantly smaller than the angular resolution capabilities of our instruments, hence in all cases, we modelled the star as a point source. 

%The stellar radius of \hd is estimated to be $0.06$mas \citep{2021stellarrad} 

% We parameterise the geometry of the disc with two geometries -- either as an elliptical Gaussian or a ring. Our data probes mainly the first lobe of the visibility function that is only weakly dependent on the geometry of the emitting object. Therefore, we limit our analysis to these two geometries. The Gaussian profile can be characterised solely with the FWHM, inclination, and position angle. Whereas the choice of the ring model, on the other hand, is physically motivated by the fact that the SED indicates the presence of hot dust near the sublimation rim, which should result in ring-like emission geometry.
% We then fit two different ring models; thin, and variable. The thin ring has the outer diameter fixed as $1.2$ times the inner diameter, whilst the variable ring has both the inner and outer diameter set as free parameters. %The data were also split into epochs to check that inclination and position angles were well constrained. Due to the small amount of data, one the epochs are split into years and different instruments. %do i run this again but with the mircx 2021 data in each day to check for any dynamical changes? 
%Four different scenarios were considered, two of which consisted of a single stellar component, and the other two included a binary component. The emission from the circumstellar disc is modelled as a Gaussian component or ring component, referred to as 'G' and 'R' in Tables \ref{tab:achromatic} and \ref{tab:chromatic}, respectively.
We considered three sets of models: a 2D elliptical Gaussian model and two ring models with different sets of constraints (see below). Any further complex models would not have been sufficiently constrained by our data as we only probe the first lobe of our visibility data. As the closure phase signals are $\simeq0^\circ$, we only fit the squared visibility data, and any offsets are set to zero. The emission from the circumstellar material is modelled as a 2D elliptical Gaussian component, a fixed thin ring component, and a ring where the inner and outer radius are free parameters. Both Gaussian and ring models provide an inclination and position angle of the inner disc, as well as the fluxes of the system. Where these models vary is on how they prescribe the size of the object - the Gaussian models produce a FWHM whilst the ring models assume an inner gaseous cavity and provide us with an inner ring radius. We consider both achromatic and chromatic models. These are referred to in Tables \ref{tab:achromatic} and \ref{tab:chromatic} as `G', `Thin R' and `R', respectively. An arbitrary radial intensity profile for the ring models is set at $R^{-2}$.
% We also considered a binary instead of a single star are marked `B', with a fixed binary (x$=6.3$mas,y$=15.1$mas) position obtained from a grid search. 
All models were fitted for an inclination angle, $i$, and position angle, P.A., which are free parameters. Here we define a face-on inclination as $i=0^\circ$ and the position angle is defined as East from North. The visibility function shows a slight drop below 1 at the shortest available baseline ($\lambda/B=31$ mas), therefore to fit the model to the data more precisely we need emission on scales of $10$s of au, which most likely represent scattered light from the extended disc. Thus, a large halo component was included in all models. The size of this extended component has been fixed to $1000$mas to represent homogeneous emission that fills the full field of view. Keeping the inner parameters fixed, we found that a $6\%$ flux contribution for this halo component fit our overall visibility curve. We then kept this parameter fixed when fitting the inner geometry as global minimisation fails when letting it vary with all other parameters. In the achromatic models, $f_{\mathrm{disc}}$ is the relative flux for the disc. The total flux $f_{\mathrm{tot}}$ is given by $f_*+f_{\mathrm{disc}}+f_{\mathrm{halo}}$. The outer radius of the inner disc is set to be $20\%$ larger than its inner radius, for consistency with \cite{monnier2005}, whilst the full-width-at-half-maximum (FWHM) for Gaussian models is a free parameter.
%Including a halo component decreased the $\chi^2$ consistently over all models, as well as enabled the model to fit the visibility curves better. \\

% In Tables \ref{tab:achromatic} and \ref{tab:chromatic}, the models are represented by G (Gaussian fit), R (ring model), B+G for a Gaussian fit with a binary companion, and B+R for a ring model with a binary companion.

% A secondary cause for the chromaticity is because the brightness distribution changes with wavelength. At shorter H-band wavelengths, the disc may look more compact than at longer wavelengths \citep{Kluska2014A&A...564A..80K}. 
%To account for chromaticity, each model was also modelled with wavelength dependence, 
For the chromatic models, the flux density ($F_\lambda$) of the components was defined by the disc spectrum power law ($F_D$) 
\begin{equation}
\label{eq:flux}
    F_D = \mathrm{F_\mathrm{Disc}}\cdot\Bigg(\frac{\lambda}{1.65\mu m}\Bigg)^\alpha,
\end{equation}
where $\mathrm{F_\mathrm{Disc}}$ is the scale flux of the disc ($F_D/F_S$ at $1.65\mu$m), $\alpha$ is the power law component, $\lambda$ is the wavelength and $1.65\mu$m is the central wavelength in the H-band. The star spectrum ($F_S$) with relative fluxes is defined as 
\begin{equation}
    \label{eq:starflux}
    %\zeta_S\propto\lambda^{-4}.
    F_S = F_{\mathrm{S,o}} \cdot\Bigg(\frac{\lambda}{1.65\mu m}\Bigg)^{-4},
\end{equation}
%The total flux $f_{\rm{tot}}$ is given by $f_1+f_2+f_3$.
where $F_{S,o}$ is the stellar-to-total flux ratio at $1.65\mu$m.
The models were kept as simple as possible to avoid over-fitting.
% Refer back to them as e.g. equation~(\ref{eq:quadratic}).
Results for our model fits can be found in Table \ref{tab:achromatic} and Table \ref{tab:chromatic}, and the synthetic images for the chromatic models can be found in \Cref{fig:chrom_models}. The fits are discussed below.
PMOIRED uses a $\chi^2$ minimisation algorithm and the uncertainties associated with the model parameters are from bootstrapping the initial PMOIRED model fit, with the number of fits $\mathrm{N}=250$.

\begin{figure*}
    \centering
    \includegraphics[width=170mm]{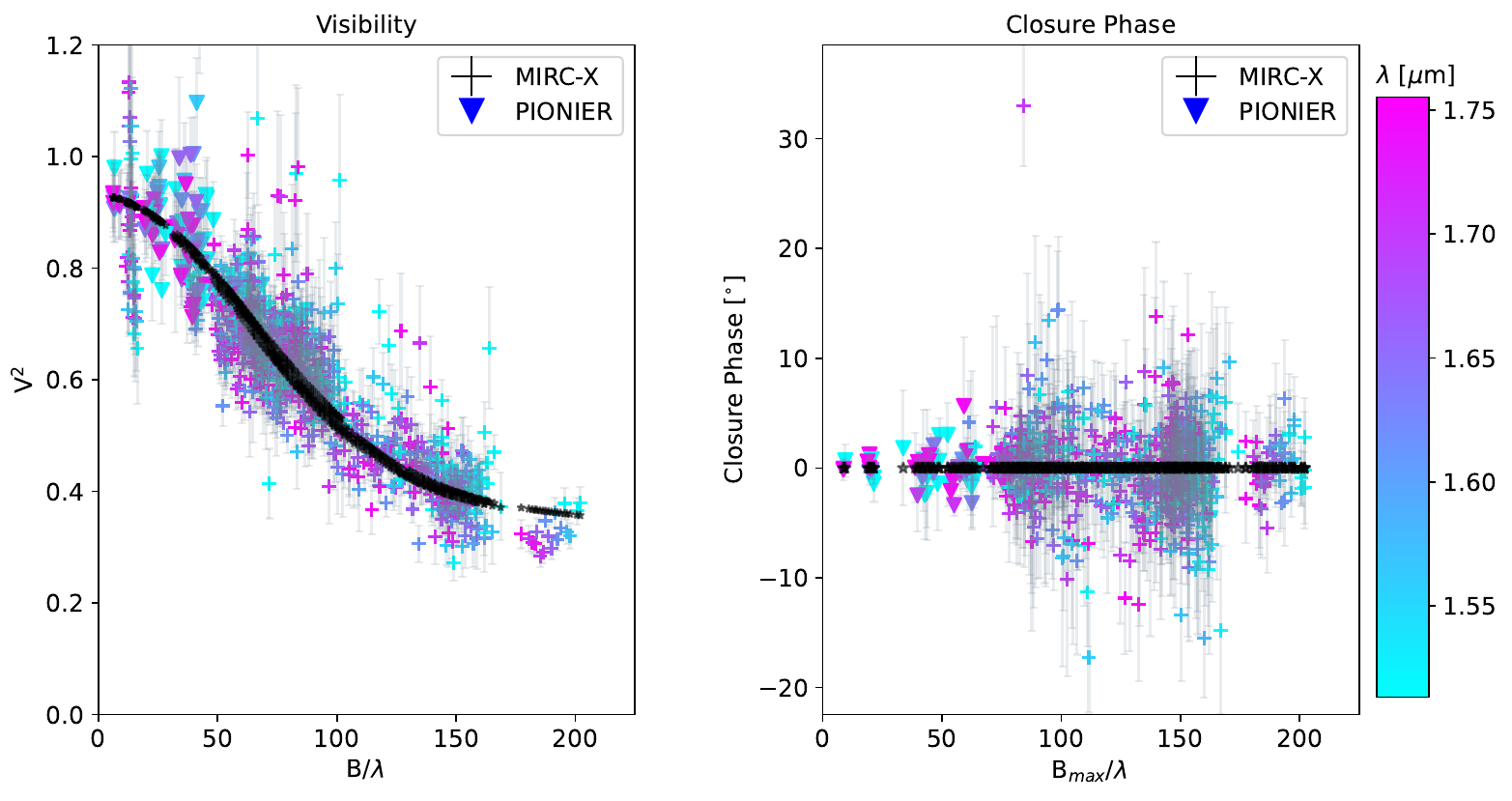}
    \caption{Achromatic combined squared visibility profile (left) and closure phase (T3PHI) from both instruments. The colour points show the measurements, while the black dots correspond to the R ring model.}
    \label{fig:ringachrom}
\end{figure*}
% \begin{figure*}
%     \centering
%     \includegraphics[width=150mm]{Images/binary+ring+chromatic.png}
%     \caption{Chromatic closure phase (T3PHI) and visibility profile fit for a binary with ring model. The {u,v} coverage is shown in the far left box.}
%     \label{fig:binaryringchromatic}
% \end{figure*}
% \begin{figure}
%     \centering
%     \includegraphics[width=85mm]{Images/uv.pdf}
%     \caption{(u,v) coverage of the combined data for all epochs on both instruments.}
%     \label{fig:UVcover}
% \end{figure}
\begin{figure*}
    \centering
    \includegraphics[width=170mm]{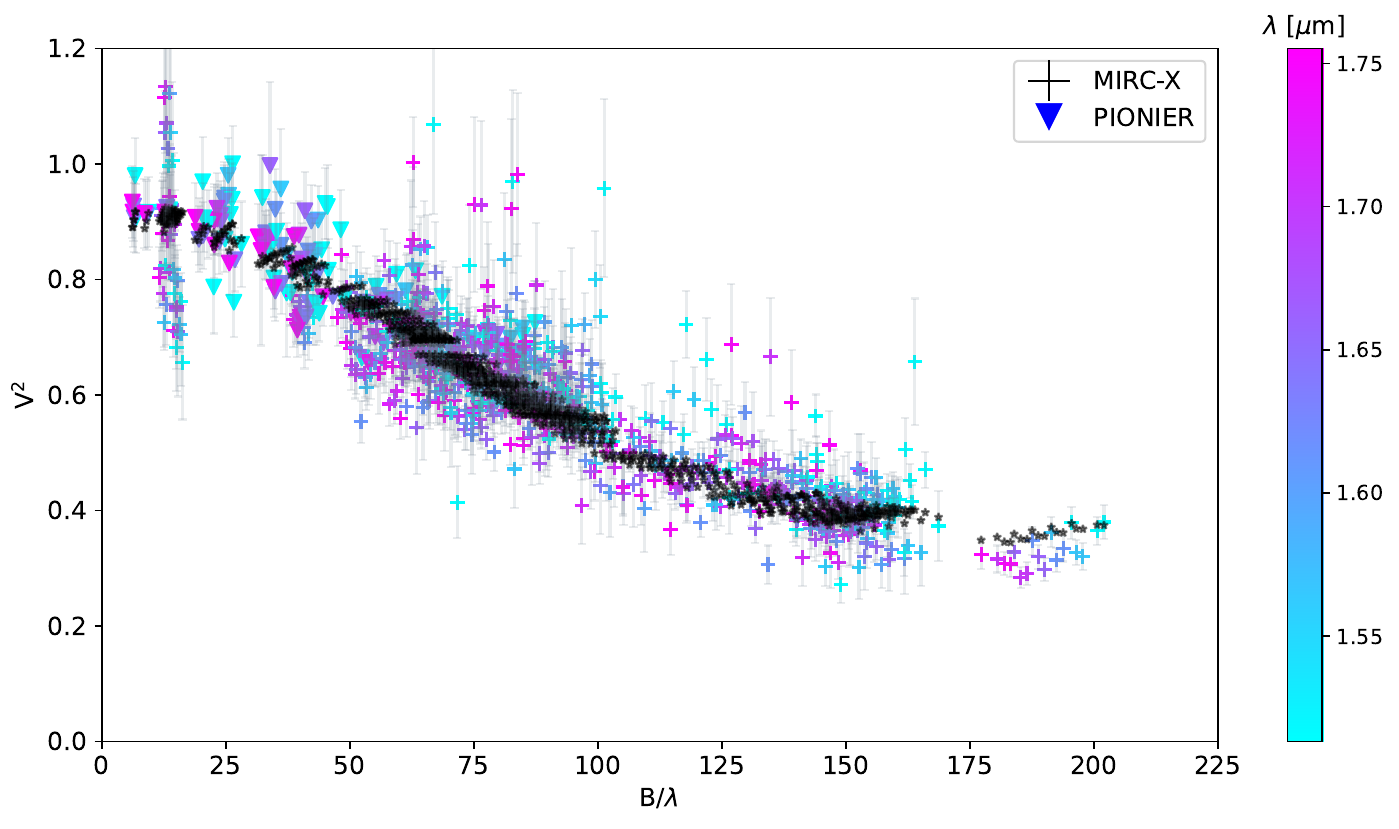}
    \caption{Chromatic combined squared visibilities from both instruments. The colour points show the data measurements, whilst the black points correspond to the R ring model.}
    \label{fig:ringchromaticVISCP}
\end{figure*}

\subsection{Misalignment and Shadows}
\label{shadowmodelling}

To calculate the misalignment angle, $\Delta\theta$, between the inner disc and outer disc, we take the inclination angle and position angle from our best-fit model, to be inputs in
% The inclination angle and position angle of the inner disc, presented in Sect. \ref{sec:incandpa} allow us to calculate the mutual misalignment angle between the inner disc and outer disc, $\Delta\theta$, with
% %This misalignment angle $\Delta\theta$ between the inner and outer disc (represented by subscript ``$1$'' and ``$2$'') is calculated using
\begin{multline}
    \label{eq:misalignment}    \cos \Delta\theta(i_{\mathrm{1}},\mathrm{PA_{1}},i_{\mathrm{2}},\mathrm{PA_{2}})=\\ \sin(i_{\mathrm{1}})\sin(i_{\mathrm{2}})\cos(\mathrm{PA_{1}}-\mathrm{PA_{2}})+\cos(i_\mathrm{1})\cos(i_\mathrm{2}),
\end{multline}
where $\mathrm{PA}$ and $i$ are the position angles and inclination angles, and where subscripts `1' and `2' refer to the inner and outer disc, respectively \citep{min2017A&A...604L..10M}. This misalignment angle is given as the two normal vectors defined by the inner and outer disc planes. As the absolute disc orientation is not known, two misalignment angles are calculated to account for which side of the inner disc is closer to the observer. These two misalignment angles are calculated as
\begin{equation}
\label{eq:innerouter}
    \Delta\theta_1=\Delta\theta(i_\mathrm{1},\mathrm{PA_{1}}, i_\mathrm{2},\mathrm{PA_{2}})
\end{equation}
and 
\begin{equation}
\label{eq:innerouter180}
    \Delta\theta_2=\Delta\theta(i_\mathrm{1},\mathrm{PA_{1}}+180^\circ, i_\mathrm{2},\mathrm{PA_{2}}).
\end{equation}
We want to determine the position angle connecting the shadows from the intersection line of the two planes of the inner and outer discs. This position angle $\alpha$ is given by \citet{min2017A&A...604L..10M}
\begin{multline}
\label{eq:PA}
    \alpha=\tan^{-1} \Bigg(\frac{\sin(i_\mathrm{1})\cos(i_\mathrm{2})\sin(\mathrm{PA_{1}})-\cos(i_{\mathrm{1}})\sin(i_{\mathrm{2}})\sin(\mathrm{PA_{2}})}{\sin(i_\mathrm{1})\cos(i_\mathrm{2})\sin(\mathrm{PA_{1}})-\cos(i_{\mathrm{1}})\sin(i_{\mathrm{2}})\cos(\mathrm{PA_{2}})}\Bigg).
\end{multline}
Equation \eqref{eq:PA} gives two sets of results due to needing to take into account whether the near sides of the inner and outer disc coincide (using Equations \eqref{eq:innerouter} and \eqref{eq:innerouter180}). In addition, we can calculate the offset of the line of shadow with respect to the central star. In the northwards direction, this is
\begin{equation}
\label{eq:shadowoffset}
    x=\frac{h\cdot\cos(i_\mathrm{1})}{\cos(i_\mathrm{2})\sin(i_\mathrm{1})\sin(\mathrm{PA_{1}})-\cos(i_{\mathrm{1}})\sin(i_{\mathrm{2}})\sin(\mathrm{PA_{2}})}.
\end{equation}

In the eastwards direction this becomes
\begin{equation}
\label{eq:shadowoffseteast}
    y=\frac{h\cdot\cos(i_\mathrm{1})}{\cos(i_\mathrm{1})\sin(i_\mathrm{2})\cos(\mathrm{PA_{2}})-\cos(i_{\mathrm{2}})\sin(i_{\mathrm{1}})\sin(\mathrm{PA_{1}})}.
\end{equation}
We define $h$ as the height of the scattering surface of the outer disc. For this paper, we vary $\frac{h}{R}$ between $5\%$ and $20\%$, where $R$ is the radius of the scattered light seen in the outer disc. For this work, we take $R=18$\,au, the radius of the disc ring where the shadows are observed \citep{perez2018ApJ...869L..50P}. 

We have then further calculated the minimum offset (in au) by
\begin{equation}
\label{eq:min_offset}
    \eta = \frac{|xy|}{\sqrt{x^2+y^2}}.
\end{equation}
% with full derivation found in \cite{min2017A&A...604L..10M}.
Fig. \ref{fig:artistrend} shows how a greater $h$ value leads to a greater offset. Fig. \ref{fig:multiscaleheight} shows a sketch of the disc geometry and how the line connecting the shadows depends on $h$. The sketches in Fig.~\ref{fig:4cases} show how the different disc orientations affect the shadow position, where the grey area indicates the observed shadow directions.
% To determine the correct orientation of the inner and outer discs, an analytical sketch was drawn for the four possible combinations of disc near-sides facing east or west with respect to the predicted shadow positions.
% discuss the different scenarios: which ones can you rule out, and which ones are consistent with all constraints?
The inclination angle and position angle used for these sketches were from the best-fit chromatic ring R model values. 
% The four combinations can be seen in Figure~\ref{fig:blenderpics}. 

\section{Results and discussion}
\label{sec:results}

\begin{table*}
%    \begin{adjustwidth}{}{}
	\centering

	%\captionsetup{justification=centering}
	\caption{HD143006 Geometric Models - Achromatic}
	\label{tab:achromatic}
	\begin{tabular}{lcccccccccr} % four columns, alignment for each
		\hline
		Model &  $i$ & $\mbox{P.A.}$ &  $f_{\mbox{disc}}$  &  $f_{\mbox{star}}$ &$f_{\mbox{halo}}$&Inner Radius & Outer Radius & FWHM &$\chi^2_{r,\mbox{vis}}$ &$\Delta\rm{BIC}$ \\
		&[$^\circ$] & [$^\circ$] & & & &(mas)&(mas) &(mas)\\
		\hline
		G & $21 \pm 3$ & $154 \pm 9$ & $0.370 \pm 0.008$ & $0.595\pm0.007$ & $0.0357$ & -&- & $1.15 \pm 0.03$& $1.72$& REF\\
		Thin R & $25 \pm 3$ & $161 \pm 6$ & $0.247 \pm 0.006$ & $0.710\pm0.006$ & $0.0426$ & $0.72 \pm 0.02$ && - & $1.80$& $99$\\
        R & $21 \pm 4$& $154 \pm10$ &$0.360 \pm0.022$ & $0.604\pm0.021$ & $0.036$ &$0.25 \pm0.04$ & $1.27 \pm0.04$& - &$1.73$ & $33$\\
		\hline
	\end{tabular}\\
    Notes: The $\chi^2_r$ for the closure phase is 0.92 for all models. As the closure phase signal is around zero, no azimuthal asymmetry was considered.\\
    A negative $\Delta$BIC signifies a better fit ($\Delta\mathrm{BIC}=\mathrm{BIC} - \mathrm{BIC_{reference\,model}})$. 
	% \\[10pt] The models run were a Gaussian fit (G), ring model (R), a Gaussian with a binary companion (B+G) and a ring model with a binary companion (B+R). All fits were run with a fixed binary (x$=1.7$mas,y$=3.5$mas) position obtained from a grid search. $f_{\mathrm{disc}}$ and $f_{\mathrm{comp}}$ are the relative fluxes for the disc and stellar companion.
%    \end{adjustwidth}
\end{table*}

\begin{table*}
	\centering
	%\captionsetup{justification=centering}
	\caption{HD143006 Geometric Models - Chromatic}
    %\begin{adjustwidth}{}{}

	\label{tab:chromatic}
    \hspace*{-1.cm}
	\begin{tabular}{lccccccccccr} % four columns, alignment for each
		\hline
		Model &  $i$ & $\mbox{P.A.}$ &  $F_{\mathrm{Disc}}$ & $\alpha$  &$F_{\mathrm{Star}}$&$F_{\mathrm{Halo}}$ & Inner Radius & Outer Radius & FWHM & $\chi^2_{r,\mbox{vis}}$& $\Delta\rm{BIC}$ \\
		&[$^\circ$] & [$^\circ$] & &&& &(mas)& (mas) &(mas) \\
		\hline
		G & $21 \pm 3$ & $155 \pm 8$ & $0.378 \pm 0.008$ & $-2.77 \pm0.21$ & $0.587\pm0.008$ & $0.035$ &- &-&$1.15 \pm 0.03$& $1.65$& REF\\
		Thin R & $26 \pm 3$ & $161 \pm 6$ & $0.254 \pm 0.006$ & $-3.06 \pm0.22$ & $0.704\pm0.006$ & $0.0423$ & $0.71 \pm 0.01$ && - & $1.75$& $113$\\
        R & $22 \pm3$ & $158 \pm8$ & $0.361\pm0.022$ & $-2.86 \pm0.25$ & $0.602\pm0.021$ & $0.036$ &  $0.26 \pm0.04$ & $1.24 \pm0.04$ & - & $1.62$& $-6.5$\\

		\hline
	\end{tabular}\\
    Notes: See Eq.\ref{eq:flux} for $F_\mathrm{Disc}$ and $\alpha$ definitions. The $\chi^2_r$ for the closure phase is 0.92 for all models. As the closure phase signal is around zero, no azimuthal asymmetry was considered.
    A negative $\Delta$BIC signifies a better fit ($\Delta\mathrm{BIC}=\mathrm{BIC} - \mathrm{BIC_{reference\,model}})$. 
    % $f_{\mathrm{disc}\mathrm{scale}}$ $f_{\mathrm{disc}\mathrm{power}}$ correspond to the fluxes (F0 and F respectively) seen in Equation \ref{eq:flux}. $f_{\mathrm{comp}}$ is the relative flux for the stellar companion.
    %\end{adjustwidth}
\end{table*}

\begin{figure}
    \centering
    \includegraphics[width=0.9\columnwidth]{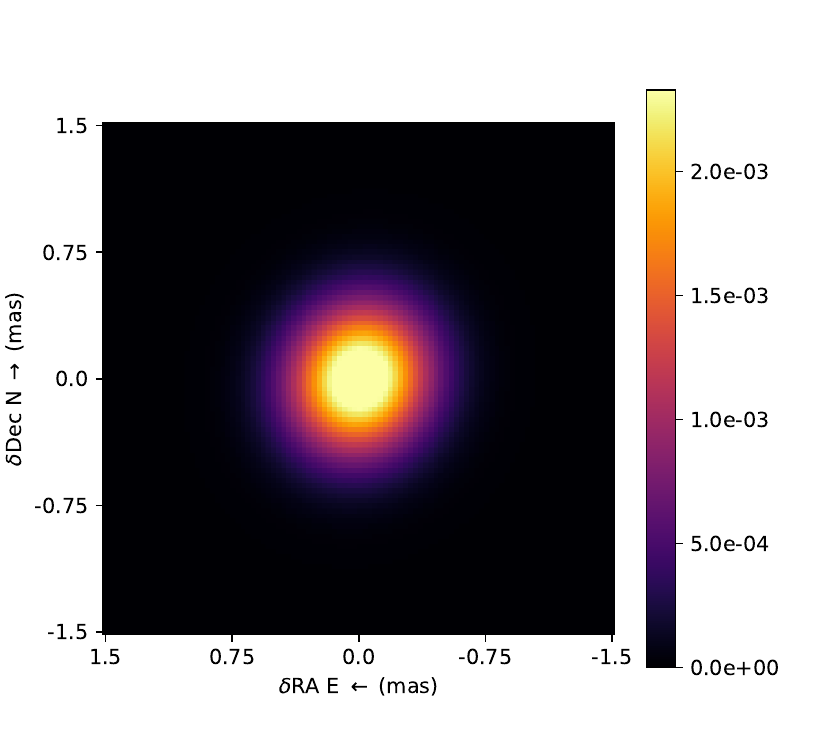}
    \vspace{-0.25cm}
    \includegraphics[width=0.9\columnwidth]{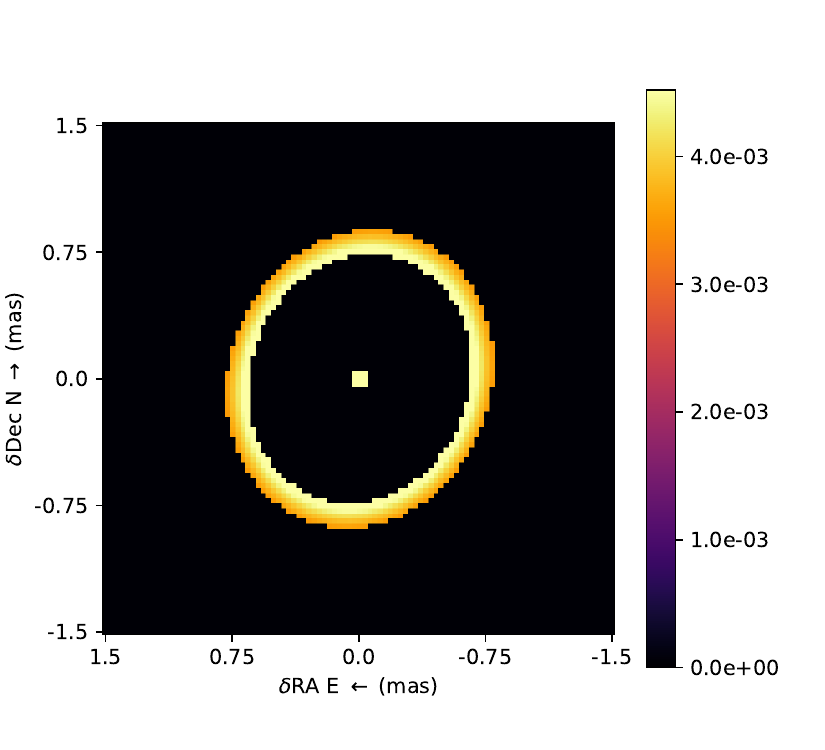} 
     \vspace{-0.25cm}
    \includegraphics[width=0.9\columnwidth]{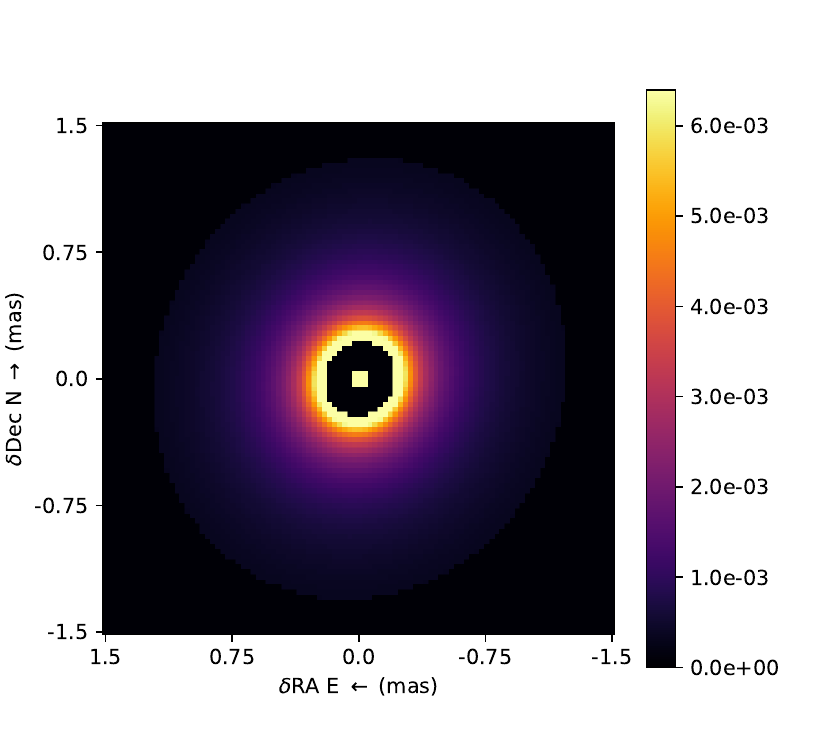}

    \caption{Best-fit chromatic models, values can be found in Table \ref{tab:chromatic}. \textit{Top:} Chromatic Gaussian model with a FWHM of $0.19$\,au. \textit{Middle:} Chromatic thin ring model with an inner rim radius of $0.12$\,au. \textit{Bottom:} Chromatic ring with variable outer radius model showing temperature dependence, with an inner rim radius of $0.043$\,au. The colour bar displays the intensity at $1.65 \mu$m, as parameterised by the best-fit models, in arbitrary units.}
    \label{fig:chrom_models}
\end{figure}

\subsection{Achromatic vs Chromatic Models}
The modelling of our data was carried out both with and without a wavelength-dependent component to investigate how much of an effect this had on the goodness of fits for the models. The inclination angles and position angles are well constrained over both the achromatic and chromatic models, whilst the reduced chi-squared ($\chi^2_r$) are consistently lower for the chromatic models (see Tables \ref{tab:achromatic} and \ref{tab:chromatic}). Whereas only a subtle difference is seen between the two, we take the chromatic models to be the more physically motivated models as evidenced by the upward arc of data points seen in Figures \ref{fig:rawdata}, \ref{fig:ringachrom} and \ref{fig:ringchromaticVISCP}.

\subsection{Inclination Angle and Position Angle}
\label{sec:incandpa}

We only consider single-star models (see Section \ref{sec:companion} for a discussion on the potential presence of a stellar/sub-stellar mass companion). We find the best-fit inclination angle and position angle from the chromatic Ring model to be $i=22.0^\circ\pm3.3^\circ$ and $\mathrm{PA}=157.6^\circ\pm8.3^\circ$. Previously, as part of a VLTI/PIONIER Herbig AeBe survey, \cite{laz2017A&A...599A..85L} observed \hd and produced an ellipsoid model and three ring models. For comparison, the inferred inclination and position angles for their ellipsoid model and ring model with no azimuthal modulation are: $i=27^\circ \pm 3^\circ$, $\mathrm{PA}=1^\circ \pm 13^\circ$ and $i=27^\circ \pm 4^\circ$, $\mathrm{PA}=180^\circ \pm 13^\circ$ respectively. All models from \cite{laz2017A&A...599A..85L} appear to fit the data equally well, as the data were only marginally resolved. Having results that strongly depend on the models used do not allow for an accurate depiction of the inner disc geometry. With the model results from this work (Tables \ref{tab:achromatic} and \ref{tab:chromatic}) we find that the inclination angles and position angles are robustly constrained and for both achromatic and chromatic, the inclinations and position angles are within a few degrees of each other. Having model-independent results allows us to more reliably take these outputs and use them as inputs for calculating the misalignment angle and shadow position (see Section \ref{sec:misalignshadows}).

% the $\chi^2_r$ values are all slightly larger, indicative of a slightly worse fit. The $\Delta$BIC for the Thin Ring model indicates a worse fit than the corresponding model in Table \ref{tab:chromatic}, however the $\Delta$BIC for the Gaussian model implies no real change in the fit, and the Ring model is ever so slightly worse. 
% The Gaussian $\chi^2_r$ in Table \ref{tab:perezchromatic} is only $0.01$ larger than in Table \ref{tab:chromatic} so the $\Delta$BIC value obtained is to be expected. There is a $0.05$ difference in the $\chi^2_r$ for the Thin Ring model so I would expect the bic to be worse. but what is happening for the ring.

\begin{table*}

	\centering
	%\captionsetup{justification=centering}
	\caption{HD143006 Geometric Models - Fixed face-on disc orientation.}
    %\begin{adjustwidth}{}{}

	\label{tab:faceon}
    %\hspace*{-1.cm}
    
	\begin{tabular}{lccccccccr} % four columns, alignment for each
		\hline
		Model &  $F_{\mathrm{Disc}}$ & $\alpha$  &$F_{\mathrm{Star}}$&$F_{\mathrm{Halo}}$ & Inner Radius & Outer Radius & FWHM & $\chi^2_{r,\mbox{vis}}$& $\Delta\rm{BIC}$ \\
		& &&& &(mas)& (mas) &(mas)&& \\
		\hline
		G  & $0.374 \pm 0.008$ & $-2.85 \pm0.23$ & $0.590\pm0.007$ & $0.035$ &- &-&$1.12 \pm 0.02$& $1.67$& $14$\\
		Thin R  & $0.252 \pm 0.006$ & $-3.14 \pm0.20$ & $0.705\pm0.006$ & $0.0423$ & $0.67 \pm 0.02$ && - & $1.82$& $65$\\
        R & $0.377\pm0.026$ & $-2.90 \pm0.28$ & $0.588\pm0.025$ & $0.035$ &  $0.22 \pm0.03$ & $1.23 \pm0.04$ & - & $1.65$& $14$\\

		\hline
        
	\end{tabular}\\
    Notes: See Eq.\ref{eq:flux} for $F_\mathrm{Disc}$ and $\alpha$ definitions. The $\chi^2_r$ for the closure phase is 0.92 for all models. As the closure phase signal is around zero, no azimuthal asymmetry was considered.
    A negative $\Delta$BIC signifies a better fit ($\Delta\mathrm{BIC}=\mathrm{BIC} - \mathrm{BIC_{reference\,model}})$. In this case, the reference model is the corresponding chromatic model from Table \ref{tab:chromatic}. 
    % $f_{\mathrm{disc}\mathrm{scale}}$ $f_{\mathrm{disc}\mathrm{power}}$ correspond to the fluxes (F0 and F respectively) seen in Equation \ref{eq:flux}. $f_{\mathrm{comp}}$ is the relative flux for the stellar companion.
    %\end{adjustwidth}
    
\end{table*}

\begin{table*}

	\centering
	%\captionsetup{justification=centering}
	\caption{HD143006 Geometric Models - Fixed inclination and position angle to outer disc values from \citet{perez2018ApJ...869L..50P}}
    %\begin{adjustwidth}{}{}

	\label{tab:perezchromatic}
    %\hspace*{-1.cm}
     
	\begin{tabular}{lccccccccr} % four columns, alignment for each
		\hline
		Model &  $F_{\mathrm{Disc}}$ & $\alpha$  &$F_{\mathrm{Star}}$&$F_{\mathrm{Halo}}$ & Inner Radius & Outer Radius & FWHM & $\chi^2_{r,\mbox{vis}}$& $\Delta\rm{BIC}$ \\
		& &&& &(mas)& (mas) &(mas)&& \\
		\hline
		G  & $0.378 \pm 0.008$ & $-2.78 \pm0.24$ & $0.587\pm0.008$ & $0.035$ &- &-&$1.14 \pm 0.02$& $1.66$& $-0.09$\\
		Thin R  & $0.252 \pm 0.007$ & $-2.98 \pm0.18$ & $0.706\pm0.007$ & $0.0423$ & $0.69 \pm 0.02$ && - & $1.77$& $72$\\
        R & $0.367\pm0.023$ & $-2.86 \pm0.25$ & $0.597\pm0.021$ & $0.036$ &  $0.24 \pm0.03$ & $1.23 \pm0.04$ & - & $1.63$& $0.22$\\

		\hline
        
	\end{tabular}\\
    Notes: See Eq.\ref{eq:flux} for $F_\mathrm{Disc}$ and $\alpha$ definitions. The $\chi^2_r$ for the closure phase is 0.92 for all models. As the closure phase signal is around zero, no azimuthal asymmetry was considered.
    A negative $\Delta$BIC signifies a better fit ($\Delta\mathrm{BIC}=\mathrm{BIC} - \mathrm{BIC_{reference\,model}})$. In this case, the reference model is the corresponding chromatic model from Table \ref{tab:chromatic}. 
    % $f_{\mathrm{disc}\mathrm{scale}}$ $f_{\mathrm{disc}\mathrm{power}}$ correspond to the fluxes (F0 and F respectively) seen in Equation \ref{eq:flux}. $f_{\mathrm{comp}}$ is the relative flux for the stellar companion.
    %\end{adjustwidth}
    
\end{table*}

% This strong model dependence of the \citep{laz2017A&A...599A..85L} results are not present here as we have a best-fit model. These results are robust and model-indepedent as all models used are within a few degrees of each other. These reliable/robust(?) results are therefore valuable for future modellings works on this object bleugh.

% PIONIER observations of \hd from a large Herbig AeBe survey \citep{laz2017A&A...599A..85L} were used by \cite{benisty2018A&A...619A.171B} to fit analytical models of the H-band emission. It was only marginally resolved and therefore all models considered fit the data equally well. The ring model with m$=1$ modulation gave $i=23^\circ\pm 5^\circ$ and $\mathrm{PA}=168^\circ\pm15^\circ$, in agreement with our results in Table \ref{tab:chromatic} where $i=24.5^\circ\pm3.4^\circ$ and $\mathrm{PA}=156.0^\circ\pm7.2^\circ$. \citet{benisty2018A&A...619A.171B} states that as only a small extent of the region was probed and the observations had a very limited angular resolution, these values are only a rough estimate of the inner disc geometry. 

% give inc and pa results, and discuss/present their robustness between models. they can be trusted and are model independent effectively. highlight that inner disc measured by lazareff was only marginally resolved therefore is only a rough estimate so this really is first time getting good measurements of it and it's super important!

\subsection{Statistical significance of models}
\label{sec:stats}

A criterion was used to check how statistically significant the drop in $\chi^2_r$ was when fitting different models, as simply adding more free parameters may lower the $\chi^2_r$ without meaning that it is a ``better'' model This was done using the Bayesian Information Criterion (BIC) value \citep{bic2015A&A...584A..72M}, computed by
\begin{equation}
    \mathrm{BIC}=-2\ln\mathcal{L}+k\ln(n),
\end{equation}
where $k$ is the number of free parameters, $n$ is the number of data points, and $\mathcal{L}$ is the likelihood function (given by $-2\ln\mathcal{L}=\chi^2$ in our models). Calculating $\Delta\mathrm{BIC}$ provides information on the statistical significance of the $\chi^2_r$ change. In both the achromatic and chromatic models (Tables \ref{tab:achromatic} and \ref{tab:chromatic}), the Gaussian model was used as the reference. Therefore a positive $\Delta\rm{BIC}$ ($\Delta\mathrm{BIC}=\mathrm{BIC} - \mathrm{BIC_{reference\,model}}$) signifies a worse fit than the Gaussian model, and a negative signifies a better fit \citep{biclimit2007MNRAS.377L..74L}. From Table \ref{tab:achromatic} we can see that the Gaussian model has the smallest $\chi^2_r$, and both the achromatic Thin R and R model have positive $\Delta$BIC values. 
In Table \ref{tab:chromatic} the chromatic ring model R fares better both in the $\chi^2_r$ and $\Delta\mathrm{BIC}$. We take this to be the best-fit model, likely indicating that the innermost ring is wide, rather than the assumption of the outer diameter being $20\%$ larger than the inner diameter. We also consider the ring models to be a more physically motivated choice due to the expectation that dust should sublimate close to the star. Similar results for wider rings or inner rings with a more Gaussian profile can be found in \cite{woitke2019PASP..131f4301W}. 

% From Table \ref{tab:achromatic} we can see that the only better model is that of the Gaussian with a binary component. However as we see no evidence for a binary detection, we do not count this result as significant. Therefore the Gaussian model remains the best fit. In Table \ref{tab:chromatic}, the ring model with the free inner and outer diameter parameter is about the same goodness of fit as the Gaussian model (again we discount the Gaussian with added binary companion). Whilst the Gaussian models remain the best fit for both achromatic and chromatic models, these are less representative of the physical process occurring at the inner sublimation rim. 
For further analysis in this paper using the disc geometry, we use the results from the chromatic ring model R.

\subsubsection{Coplanar With Outer Disc?}
Two further checks were carried out to investigate the robustness of our inclination angle result. The first of these was fixing the disc to be face on ($i=0^\circ$) and the second case fixing the inclination and position angle to that of the outer disc found in \citet{perez2018ApJ...869L..50P} ($i=17^\circ, \mathrm{PA}=176^\circ$). We only considered chromatic models, and the reported $\Delta$BIC values use the corresponding models in Table \ref{tab:chromatic} as references.

For the face-on models, the $\Delta$BIC indicate that these are worse models and the innermost disc is unlikely to be completely face-on. 
For the models where the inclination and position angle is fixed to the outer disc value from \citet{perez2018ApJ...869L..50P}, the $\chi^2$ shows that the fits are all worse, however by removing two free parameters the models with the fixed orientation perform better looking at the $\Delta$BIC as it heavily penalises models with more free parameters. For these cases in Tables \ref{tab:faceon} and \ref{tab:perezchromatic}, there are 2 less free parameters than the models in Table \ref{tab:chromatic}.
Therefore we can conclude that that whilst a face-on $i=0^\circ$ case is unlikely, the disc is also unlikely to have a larger inclination angle than what we have obtained through our modelling efforts.

\subsection{Misalignment and Shadows}
\label{sec:misalignshadows}
% discuss either here or earlier than shadow doesn't depend on radius

\begin{figure*}

\begin{subfigure}[b]{.45\linewidth}
\includegraphics[width=\linewidth]{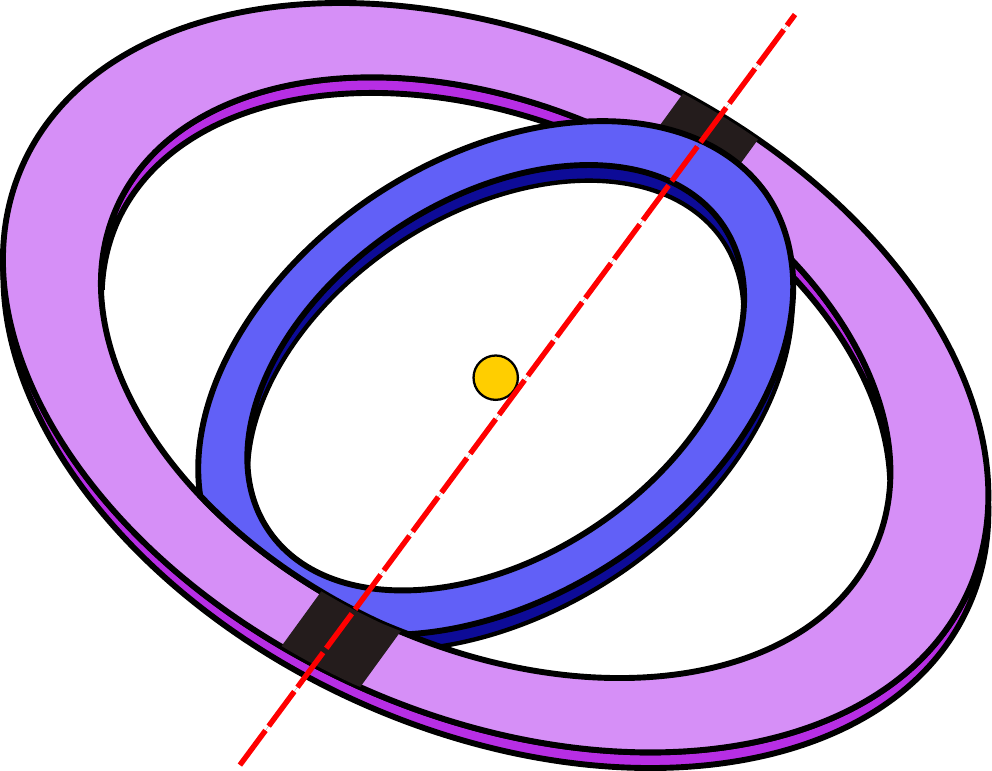}
%\caption{A gull}\label{fig:gull}
\end{subfigure}
\begin{subfigure}[b]{.45\linewidth}
\includegraphics[width=\linewidth]{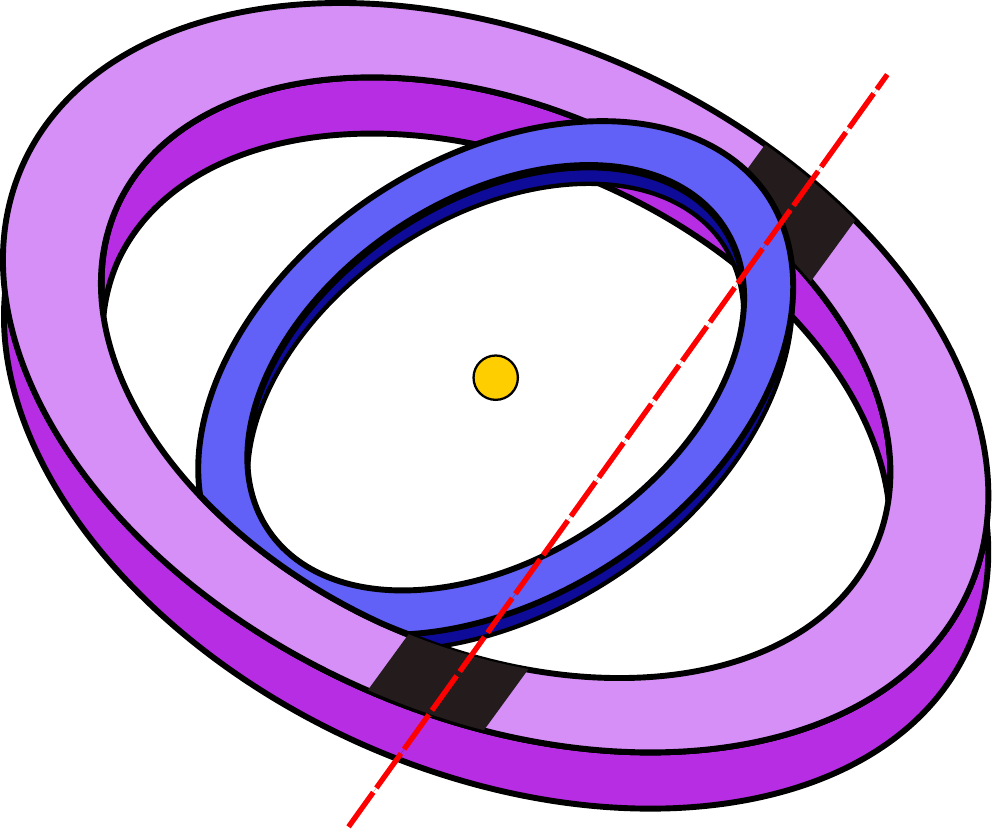}
%\caption{A tiger}\label{fig:tiger}
\end{subfigure}

\caption{Cartoon showcasing an inner and outer disc misalignment, and how the scale height of the outer disc causes the line of shadow (red dashed line) to be offset from the central star. For an infinitely thin disc, the line of shadow would appear to pass directly through the central star.}
\label{fig:artistrend}
\end{figure*}

\begin{figure}
    \centering
    \includegraphics[width=\linewidth]{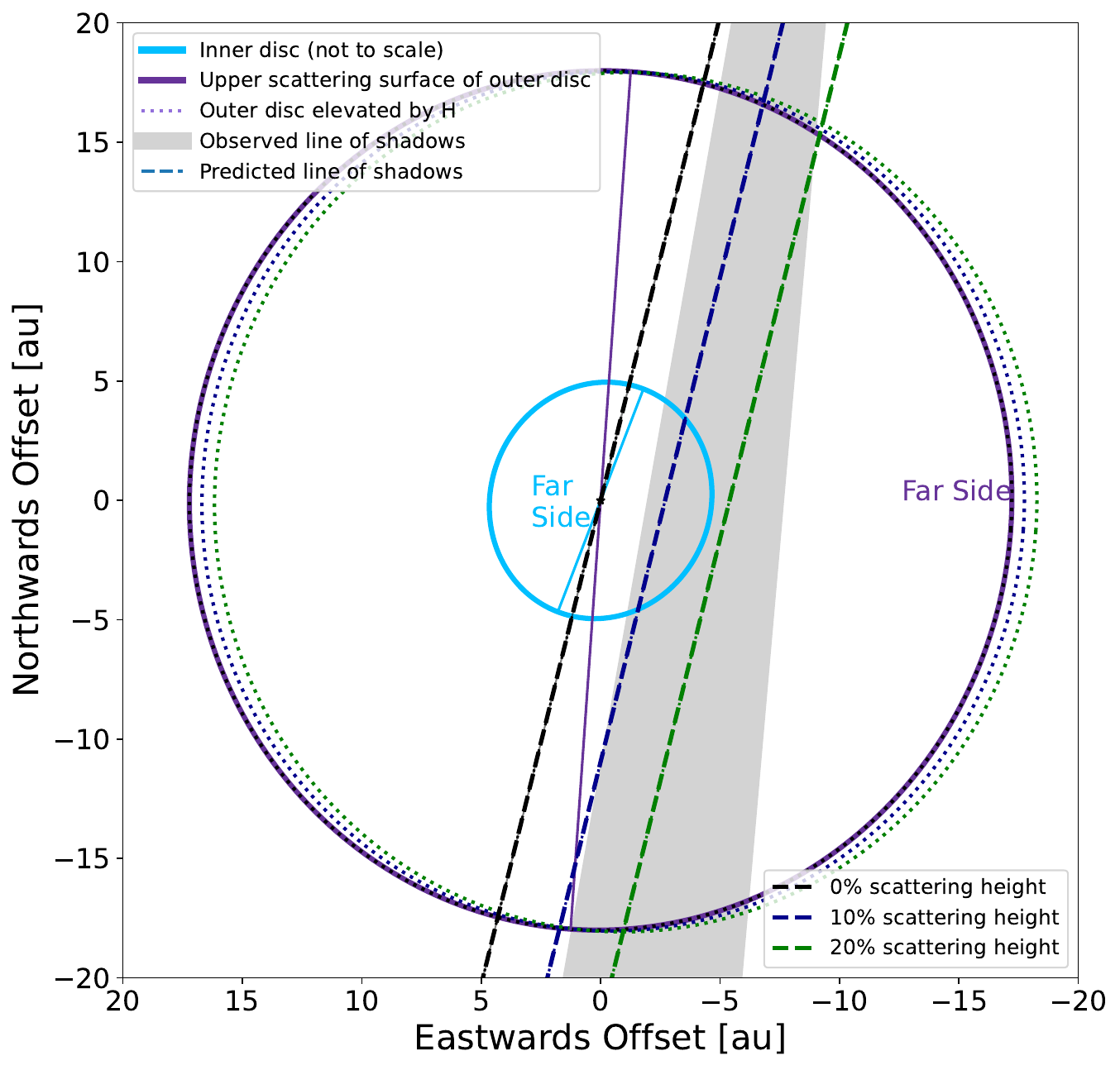}
    \caption{Analytical sketch of the outer disc (purple) and inner disc (blue, not to scale), showing how a greater scattering height percentage creates a larger offset of the line of shadow from the central star. As shown by the black dashed line, assuming a scattering height of 0 should mean the line of shadow passes through the central star. If the far sides of the inner and outer discs are swapped, the line of shadow is shifted to the left of the central star. This allows us to constrain inner/outer disc orientation.}
    \label{fig:multiscaleheight}
\end{figure}

\begin{figure*}%[tbp]
  \centering
  \begin{tabular}{c}
    %\begin{minipage}{20cm}
      $\begin{array}{cc}
        \includegraphics[width=7.2cm,angle=0]{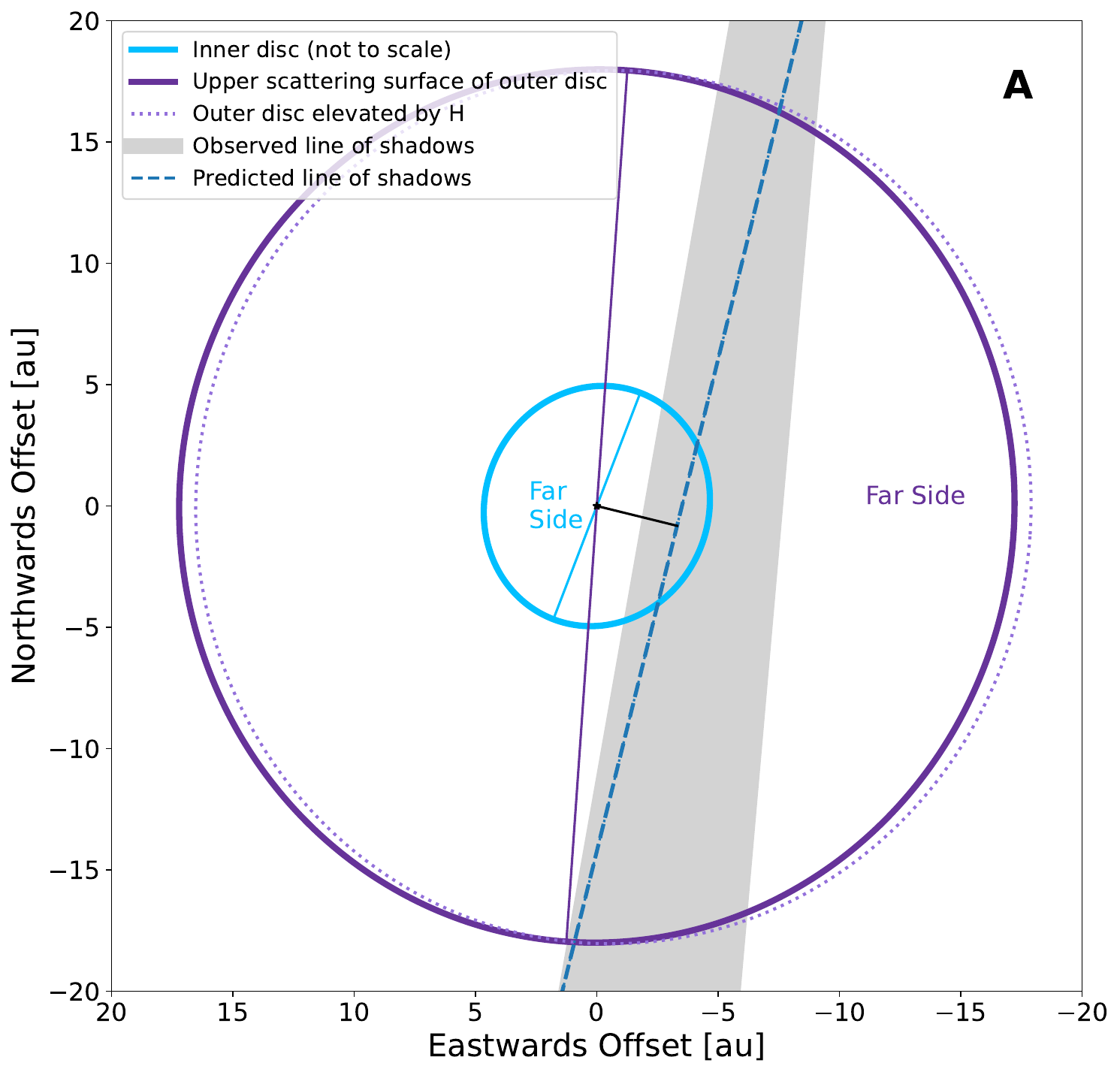}       & 
        \includegraphics[width=7.2cm,angle=0]{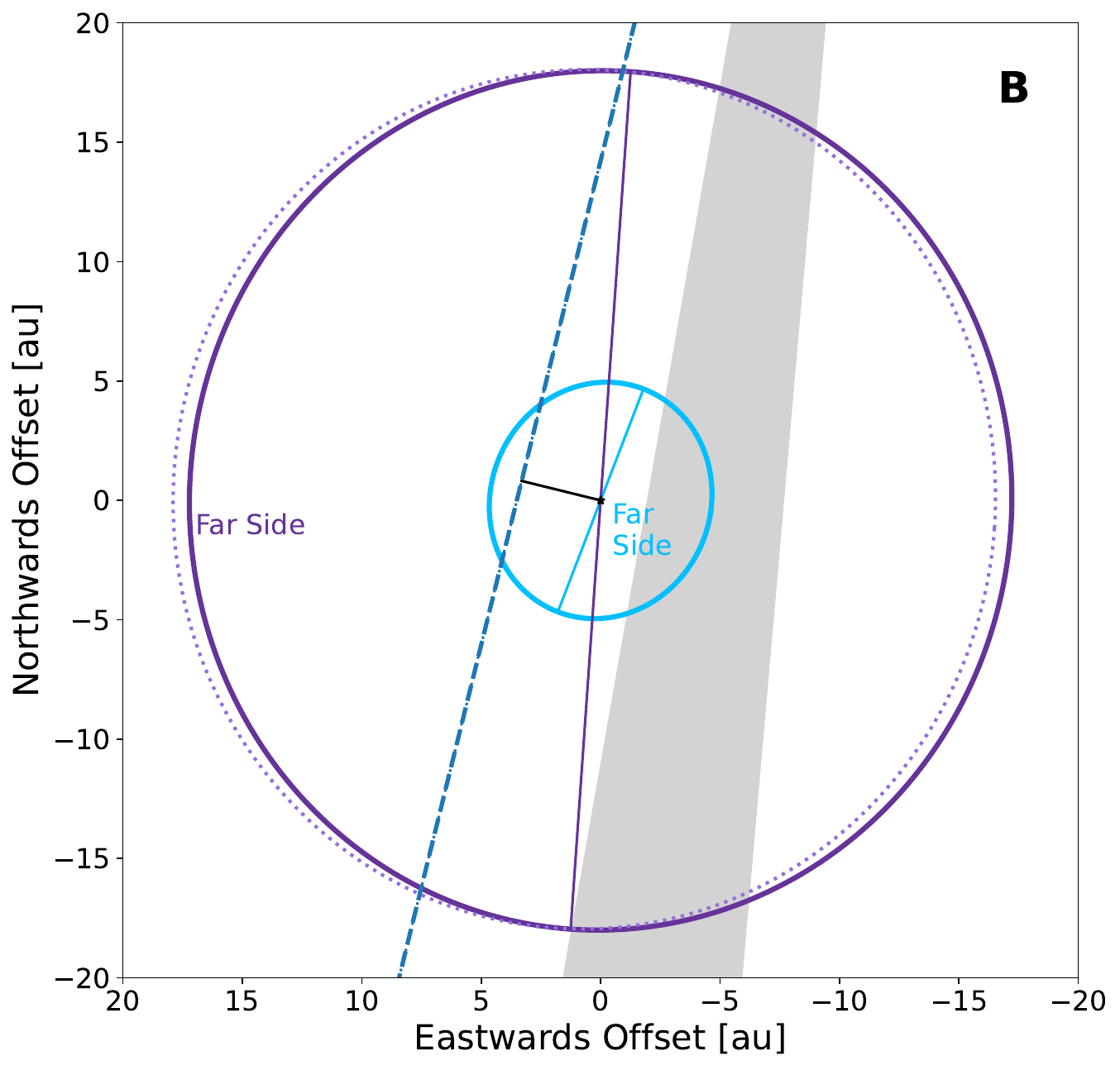} \\
        \includegraphics[width=7.2cm,angle=0]{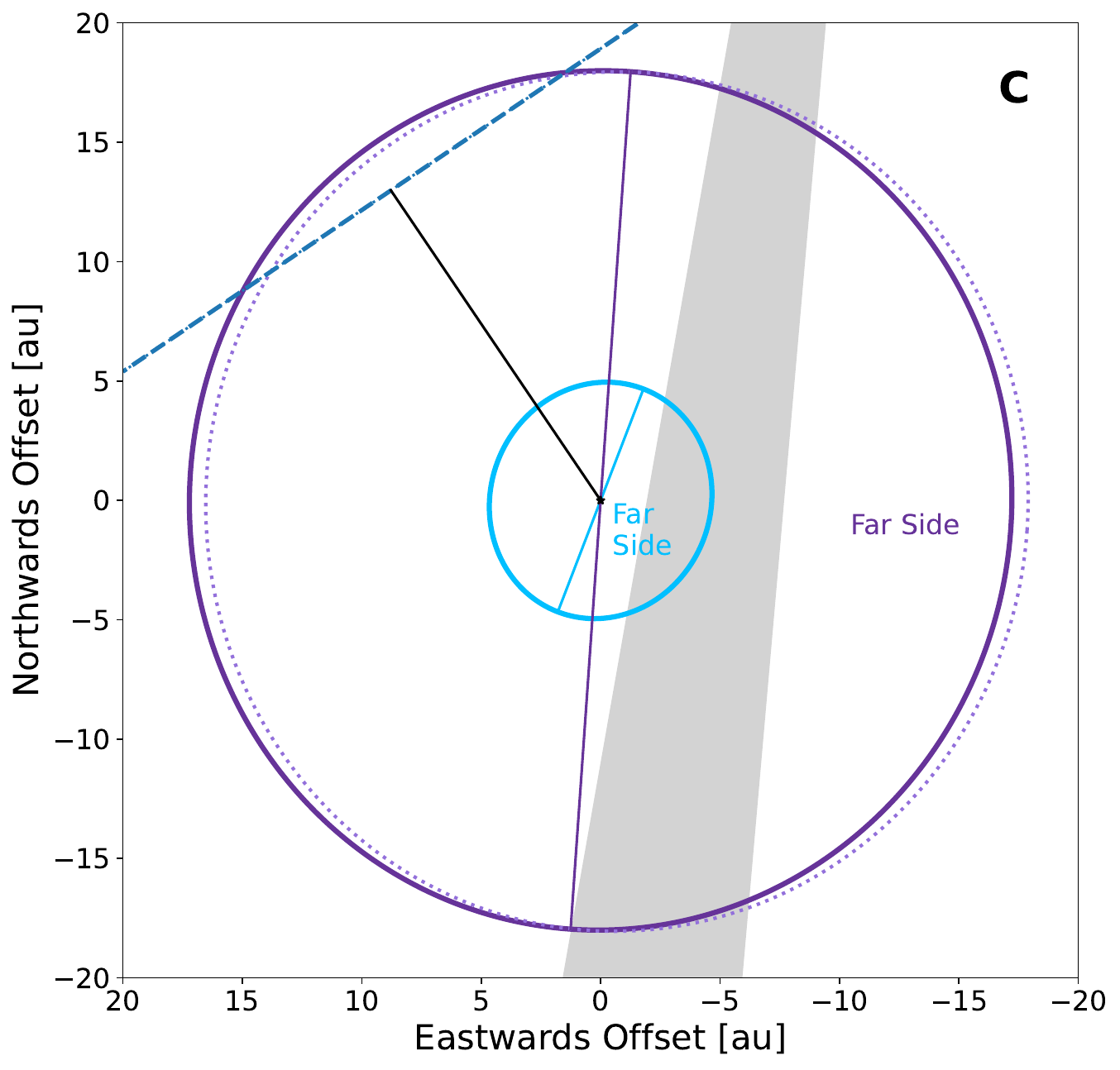}       & 
        \includegraphics[width=7.2cm,angle=0]{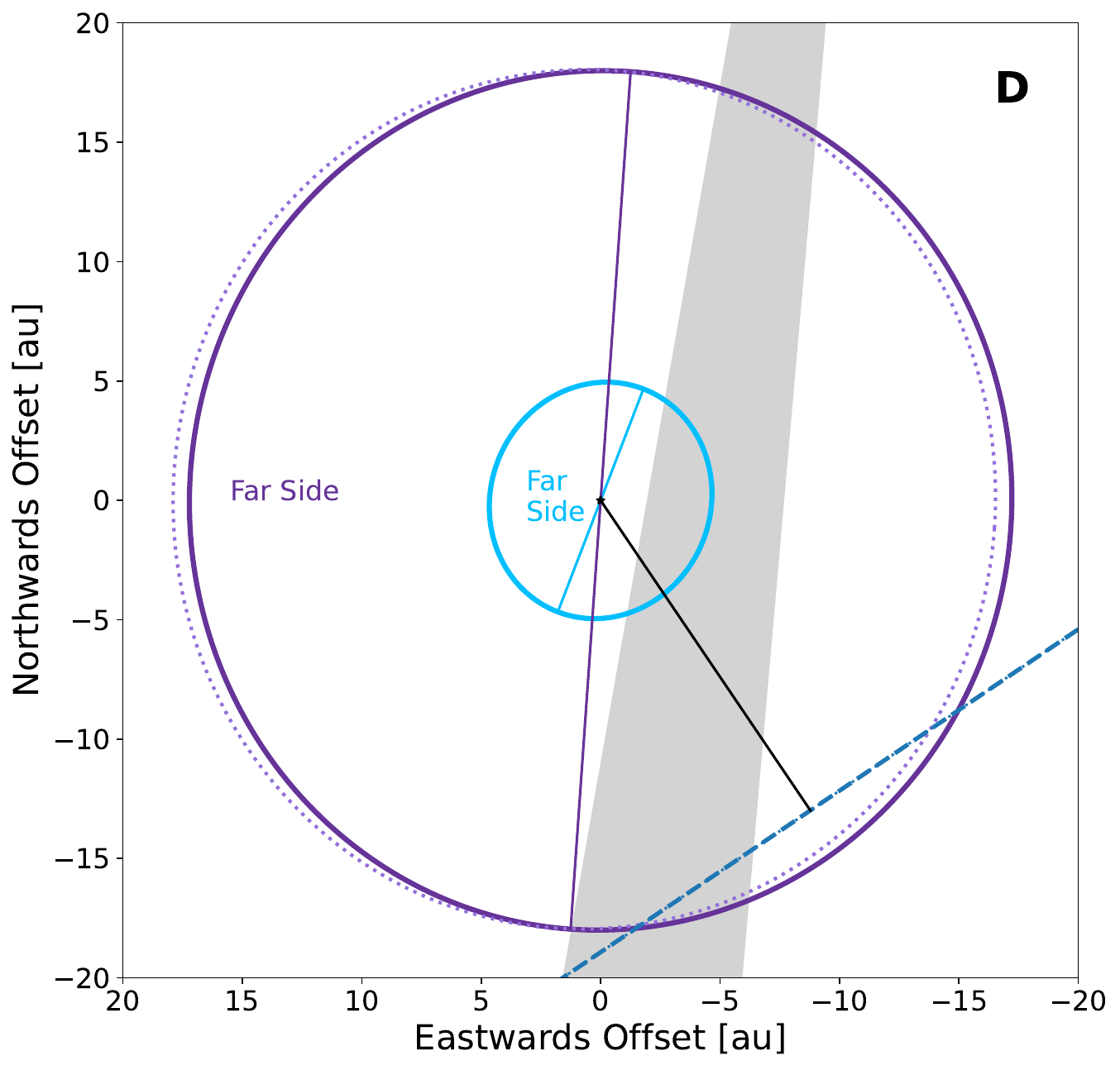} \\
      \end{array}$
    %\end{minipage}
  \end{tabular}
  \caption{Analytical sketches of all 4 cases of disc orientations, showing how the shadow positions change depending on the direction of the near side of the inner and outer disc. The dashed blue line in the analytical sketches is where we predict the shadow should lie following \citet{min2017A&A...604L..10M}, and labelled in purple and blue are the far sides of the outer and inner disc respectively. The inner disc is not to scale and is just a visual representation as the size disparity between the two discs is too large. We constrain $h/R\sim13\%$ which displaces the shadow from passing through the central star causing an offset as shown by the black line (See Figure \ref{fig:multiscaleheight}). Case A is our best-fit model as discussed in Section \ref{sec:misalignshadows}.
    }
  \label{fig:4cases}
\end{figure*}

We investigate four different cases of inner/outer disc orientation using Equations \eqref{eq:innerouter} and \eqref{eq:innerouter180} with Equation \eqref{eq:misalignment}, confirming that with our measurements of the inner disc orientation, the shadows seen in \hd can be explained by a misalignment between the inner and outer disc. Figure \ref{fig:4cases} showcases how the line of shadow (Eqns.~\ref{eq:PA}--~\ref{eq:min_offset}) changes depending on the orientation of the inner and outer disc. These four cases are as follows:
\begin{itemize}
    \item Panel A: Far side of inner disc east, far side of outer disc west,
    \item Panel B: Far side of inner disc west, far side of outer disc east,
    \item Panel C: Far side of inner disc west, far side of outer disc west,
    \item Panel D: Far side of inner disc east, far side of outer disc east.
\end{itemize}
The mutual misalignment angle is found to be $\Delta\theta_2=39^\circ\pm4^\circ$ for panels A and B and $\Delta\theta_1=8^\circ\pm4^\circ$ for panels C and D, with an apparent shadow offset of $\sim3.5$\,au and $\sim16$\,au, respectively. As per \citet{min2017A&A...604L..10M} and seen in Equations \ref{eq:shadowoffset}, \ref{eq:shadowoffseteast} and \ref{eq:min_offset}, the apparent `offset' that we see is directly linked to the scattering surface of the outer disc. As $h$ increases, so does the apparent offset of the shadow. This effect can be seen for a general case in \Cref{fig:artistrend}, and can be seen more specifically for this disc in \Cref{fig:multiscaleheight}. The position angle $\alpha$ of the shadows from Equation \ref{eq:PA} (blue dotted line in Figure \ref{fig:4cases}) are $-15^\circ\pm5^\circ$ for Panels A and B, and $-63^\circ\pm5^\circ$ for Panels C and D.

\begin{figure*}
\begin{subfigure}[b]{.45\linewidth}
\includegraphics[width=\linewidth]{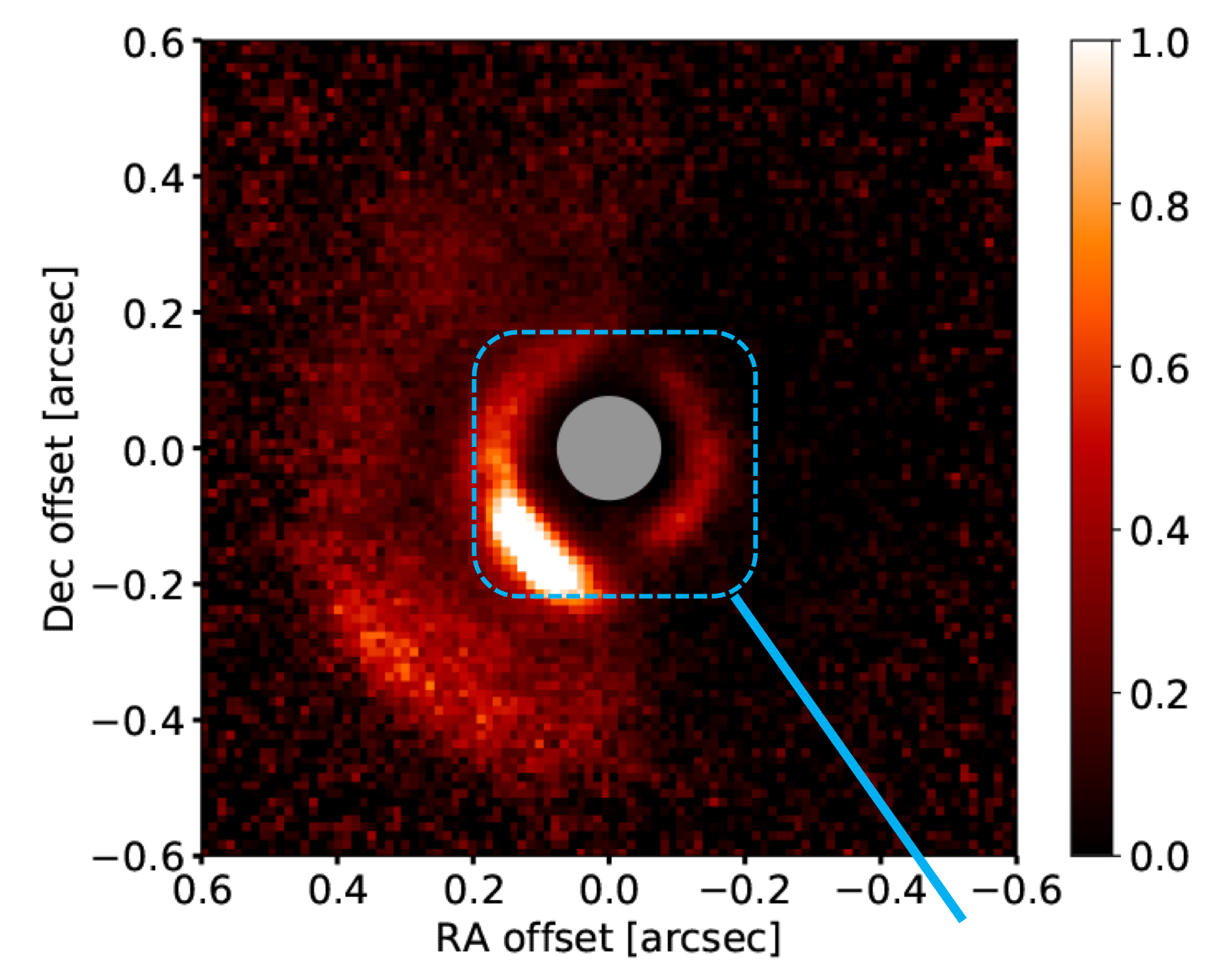}
%\caption{A gull}\label{fig:gull}
\end{subfigure}
\begin{subfigure}[b]{.45\linewidth}
\includegraphics[width=\linewidth]{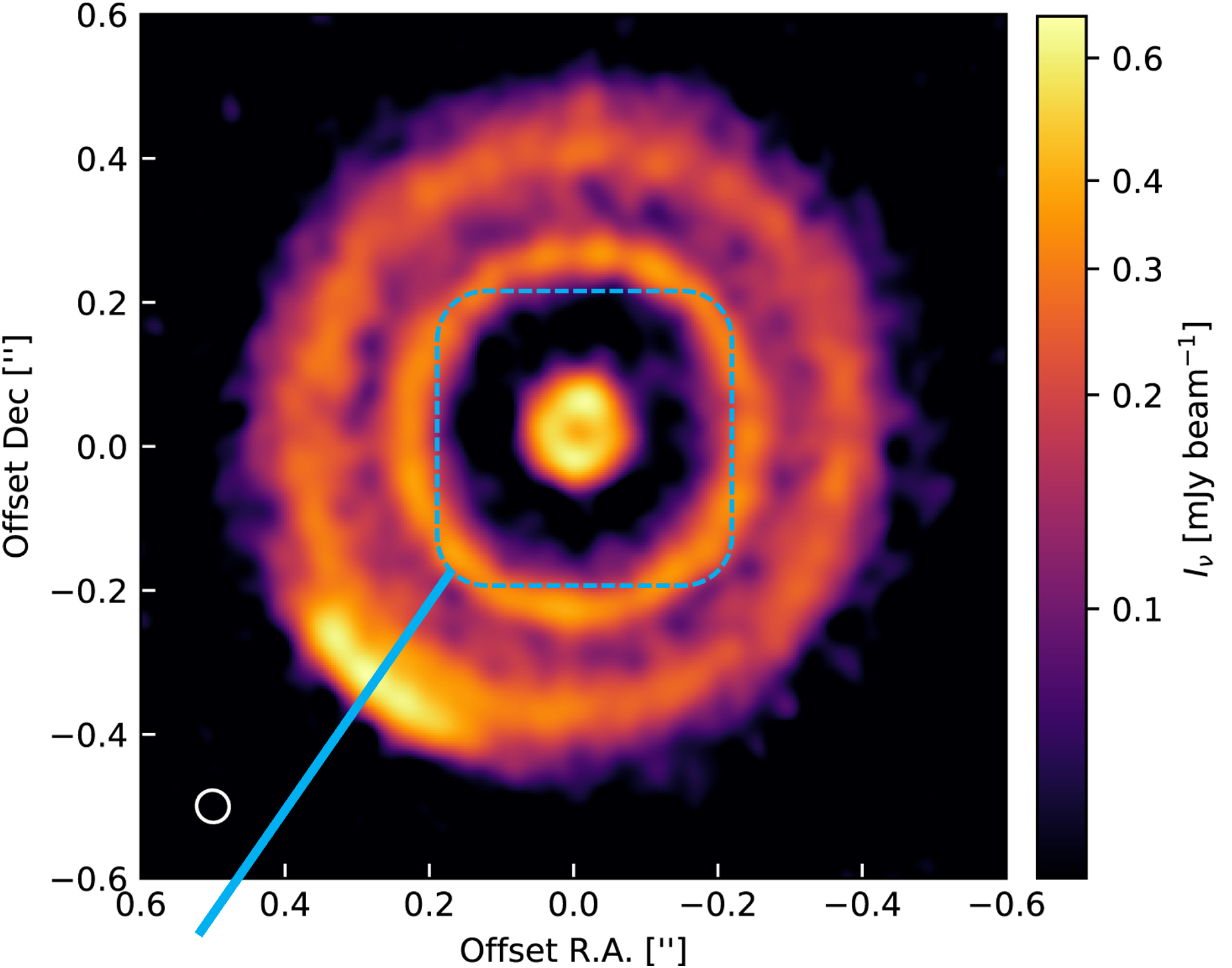}
%\caption{A tiger}\label{fig:tiger}
\end{subfigure}

\centering
\begin{subfigure}[b]{.5\linewidth}
\includegraphics[width=\linewidth]{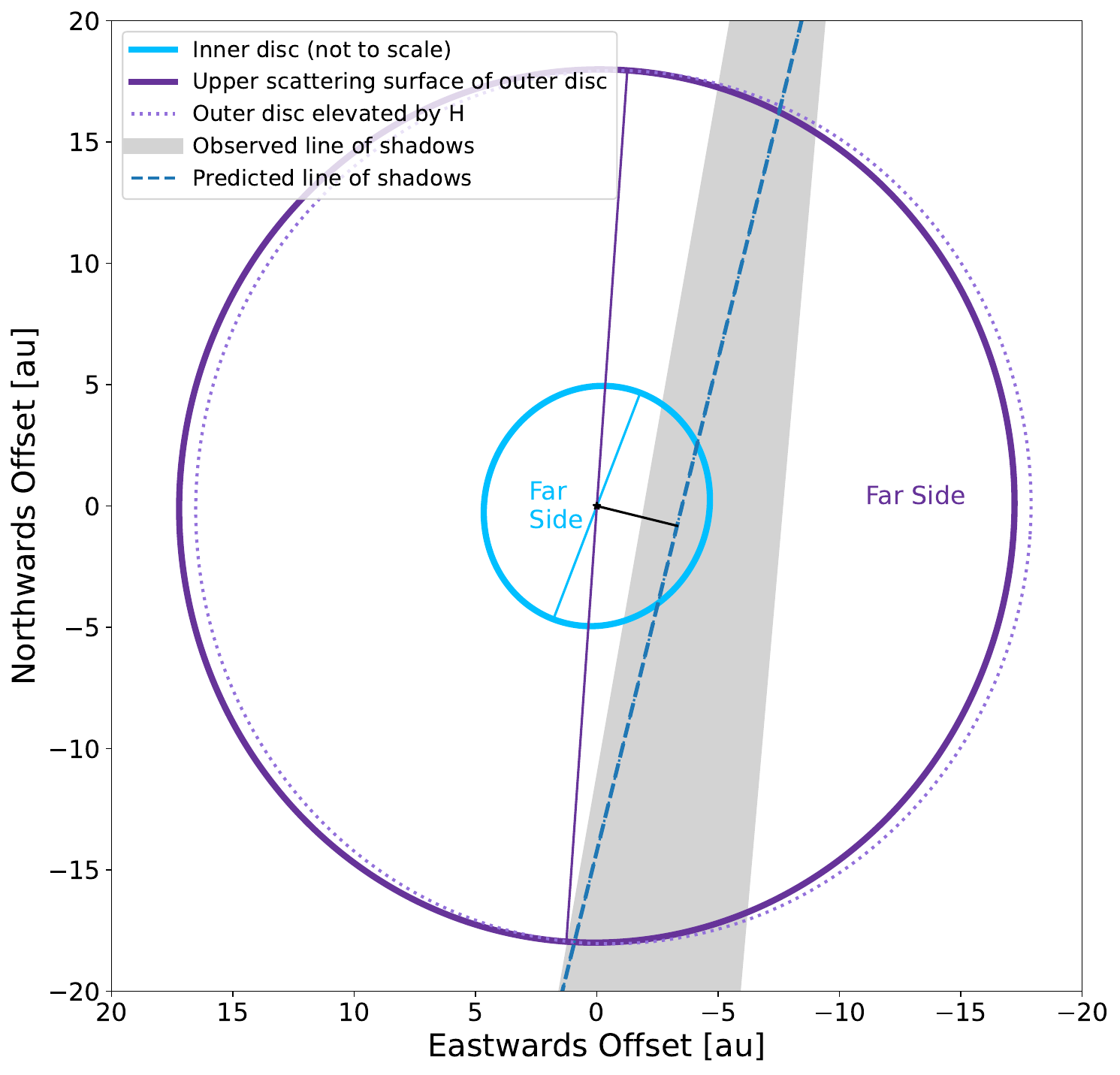}
%\caption{A mouse}\label{fig:mouse}
\end{subfigure}

\caption{\textit{Top Left:} SPHERE scattered light image from \citet{benisty2018A&A...619A.171B}. \textit{Top Right:} ALMA image from \citet{perez2018ApJ...869L..50P}. \textit{Bottom:}
Analytical sketch of where shadows lie on the outer disc due to a misaligned inner/outer disc. The radius of the inner disc has been increased for it to be visible, however this does not affect the results. The inclination angle and position angle of the inner disc are taken from this work, whilst the outer disc inclination angle and position angle are taken from \citet{perez2018ApJ...869L..50P}. The dashed blue line is where the line of shadow should appear due to observational data of the inner disc combined with the equations stated in \ref{sec:misalignshadows}. The grey shaded region are where the shadow lane appears in \citet{benisty2018A&A...619A.171B} scattered light image. Purple and blue labels indicate the direction of the far side of the outer and inner disc respectively. The solid purple circle is representative of the outer disc whilst the dashed purple line indicates some scale height value. In this case, $h/R=0.13$.}
\label{fig:analytical sketch}
\end{figure*}

\cite{benisty2018A&A...619A.171B} found that a misalignment angle of $30^\circ$ best reproduced the overall disc morphology of the shadows seen in the scattered light images. This value however was derived using strong assumptions about the inner disc orientation as it was not well resolved.
% Whilst only basic modelling parameters were used to model the inner sublimation rim of the disc, the position angle and inclination angle were well constrained. Using the best fit inclination and position angle values from the chromatic single stellar component ring model (Table \ref{tab:chromatic}) and values obtained by \citet{perez2018ApJ...869L..50P} for the outer disc, the two misalignment angles calculated were $\Delta\theta_1=8.9^\circ$ and $\Delta\theta_2=40.6^\circ$. The first misalignment angle $\Delta\theta_1=8.9^\circ$ assumes that the inner and outer discs share the same near side of the disc, the second value $\Delta\theta_2=40.6^\circ$ is calculated with the near sides not coinciding - in particular the inner near side in the west, and outer near side in the east. \\
% The misalignment angle $\theta=40.6^\circ$ obtained from the equations in Section \ref{sec:misalignshadows} derived in \cite{min2017A&A...604L..10M}, best fit with the scattered light images seen in \cite{benisty2018A&A...619A.171B} when the orientation of the inner and outer discs have their near sides opposite each other. 
% The angle of the line of shadow $\alpha$ for our best fit model is $-19.0^\circ$, matching well with the Northern shadow observed in scattered light, seen as a sketch in the bottom panel of Figure \ref{fig:analytical sketch}. 
By comparing the dotted blue line predicting the shadow locations, to the shadows found in the scattered light image by \citet{benisty2018A&A...619A.171B}, indicated in Figure \ref{fig:4cases} by the grey shaded region, we can eliminate the cases in which both the inner and outer discs have their far sides pointing in the same direction (panels C and D in Fig.\ref{fig:4cases}). The closest match between our model and the shadow position from \citet{benisty2018A&A...619A.171B} is panel A - where the inner disc has its far side facing east (also seen in Fig. \ref{fig:analytical sketch}), making our inclination $-22^\circ$.
% Our best-fit model predicts that the southern shadow should appear slightly further East than seen in the \citet{benisty2017A&A...597A..42B} image. 
% Using equation \eqref{eq:innerouter}, the misalignment angle between the inner and outer disc is $\Delta\theta_1=8.9^\circ$. Panel D in Fig. \ref{fig:4cases} depicts the case where the far side of both the inner and outer are East. This scenario gives an apparent shadow offset of $\sim14$\,au, and this value becomes negative if both far sides of the inner and outer disc are West. If we use equation \eqref{eq:innerouter180}, the misalignment becomes $\Delta\theta_2=40.6^\circ$. With the near sides facing opposite directions, there is an offset of $\sim3$\,au. As per \citet{min2017A&A...604L..10M} and seen in Equations \ref{eq:shadowoffset}, \ref{eq:shadowoffseteast} and \ref{eq:min_offset}, the apparent `offset' that we see is directly linked to the scattering surface of the outer disc. As $h$ increases, so does the apparent offset of the shadow. This can be seen in Fig.~\ref{fig:multiscaleheight} whilst the values of the offset changing with respect to $h/R$ can be seen in Table \ref{tab:scaleheights}.

Our predicted shadow position (blue dotted line, Figures \ref{fig:4cases} and \ref{fig:analytical sketch}) falling towards the bottom of the disc is almost outside of the greyed out region - this grey region in \Cref{fig:multiscaleheight,fig:4cases,fig:analytical sketch} correspond to the shadow location seen in \citet{benisty2018A&A...619A.171B}. This may be due to the bright clump (seen in Fig \ref{fig:analytical sketch} in the top left panel between declination offset $0.0$ and -$0.2$ and RA offset $0.0$ -- $0.2$) which could be causing the shadow to appear thinner, and may have a higher scattering surface. This difference is approximately within the width of the shadow itself. 
%We also note that this region of the disc contains a significantly bright asymmetry.  
% It it also possible that this bright spot could be caused by the shadow \citep{zehao2024ApJ...975..126S,ziampras2024arXiv241013932Z}. 
Another reason why the shadow position does not match the observed position towards the South could be that there appears to be some rotation of the shadow. Images of \hd published in \citet{ren2023A&A...680A.114R}, with observations taken 5 years after those in \citet{benisty2018A&A...619A.171B} show a slightly different shadow position to those in \citeauthor{benisty2018A&A...619A.171B}, which could be due to precession of the inner disc. With a majority of our data taken after the SPHERE data collection in 2016 \citep{benisty2018A&A...619A.171B} (see Table \ref{tab:hd143006obs} for observations), this could explain why there is not an exact shadow match. Analysis of this is outside of the scope of this study but would be interesting to follow up on in future works.

% \begin{table}
%     \centering
%     \begin{tabular}{cc}
%         $h/R$ & Offset \\
%         ($\%$) & (au)\\
%         \hline
%          $5$& $1.32$\\
%          $10$& $2.64$\\
%          $15$& $3.96$\\
%          $20$& $5.29$\\
%          \hline\\
%     \end{tabular}
%     \caption{Associated offsets with far side of the inner disc facing east and far side of the outer disc facing west.}
%     \label{tab:scaleheights}
% \end{table}

We can also compare our inner disc orientation with the inner disc seen in ALMA data from \citet{perez2018ApJ...869L..50P}, referred to as B8 hereafter ($\mathrm{inc}=24^\circ$, $\mathrm{PA}=164^\circ$). The inner disc in this work could be consistent with B8 if the inclination of B8 was $-24^\circ$. However \citet{perez2018ApJ...869L..50P} state that the CO kinematics favours a positive inclination and a small misalignment angle ($8^\circ$). Nevertheless, \citeauthor{perez2018ApJ...869L..50P} acknowledges that the shadow location favours a negative inclination angle and the kinematics for B8 are not highly resolved enough to exclude this orientation.
If indeed the inner disc in this work and B8 in \citet{perez2018ApJ...869L..50P} are co-planar, then there may be a warp in the disc gas between the bright ring at B8 and the intermediate ring at 40\, au (see large dark annulus in top right panel of \Cref{fig:analytical sketch}).

\subsection{Putative Companion}
\label{sec:companion}

% give results of search and also detection limits.
% but also say that the companion search i am doing is in the plane of the disc so the it's possible that the disc itself may be in the way of a companion and hiding it.

Dark regions (excluding features such as concentric gaps and rings) observed in protoplanetary discs are often attributed to shadows caused by a warped or misaligned inner disc (eg. \citealt{2015marino,benisty2017A&A...597A..42B}). These warps could be caused by companions as shown by \citet{price2018MNRAS.477.1270P} in the disc HD\,142527, with the companion also (in part) explaining the cavity, spiral arms, dust asymmetry, as well as fast radial flows seen in the gas (however some recent work with VLTI/GRAVITY by \citealt{nowak2024A&A...683A...6N} has disputed this). Evidence of gap-crossing streams has been found in AA Tau \citep{Loomis2017}, which could be perturbing the disc and causing a warp. Whilst further observations are needed to fully deduce what could cause a warp, several known transition discs have had their shadows confirmed by a misaligned inner disc, seen in \citet{bohn2022A&A...658A.183B}, with observations from GRAVITY and ALMA. 

Multiple scenarios have been proposed in which a companion can warp a disc but the underlying condition is that the angular momentum of the companion needs to be larger than the angular momentum of the inner disc \citep{Matsakos2017}. \cite{xianggruess2013} explore a range of planet masses and inclinations between $1$--$6$ $M_J$ and $10^\circ$--$80^\circ$ respectively. Their simulations find that the more massive planet with a more inclined orbit can cause a relative misalignment of up to $15^\circ$, smaller than the result obtained in this work. Conversely, \cite{Bitsch2013} find that larger disc misalignments can be created where planets have less inclined orbits. These smaller inclined planets have a stronger effect on the disc and the disc is also able to achieve an inclination angle greater than the planet. The Kozai-Lidov effect, involving a binary system with a planetary companion, can cause a wide range of misalignment angles \citep{Martin2016,Nealon2020b} however this is unlikely in our case as we do not find evidence for a stellar companion in the inner disc.
Late material infall may also cause a misalignment in the disc (eg. \citealt{Krieger2024arXiv240308388K,huang2021ApJS..257...19H,dullemond2019A&A...628A..20D,huang2023ApJ...943..107H,gupta2023A&A...670L...8G}). A further reason for the misalignment value found in this work could be due to a flyby event. \citet{Nealon2020} find inclination angles up to $45^\circ$ can occur during this events.

Pertinent to this work, the simulation carried out by \citet{Ballabio2021MNRAS.504..888B} suggest a stellar and planetary companion, with the stellar companion at an inclined orbit. The binary separation in the simulation was set to $2$ au, with this companion being $0.36 M_\odot$ (a mass ratio of $0.2$). The planetary companion was embedded at $32$ au with a mass of $10$ $M_J$. Whilst we do not have the field-of-view to search for a planetary companion, we are able to search for this putative stellar companion.

\begin{figure}
    \centering
    \includegraphics[width=90mm]{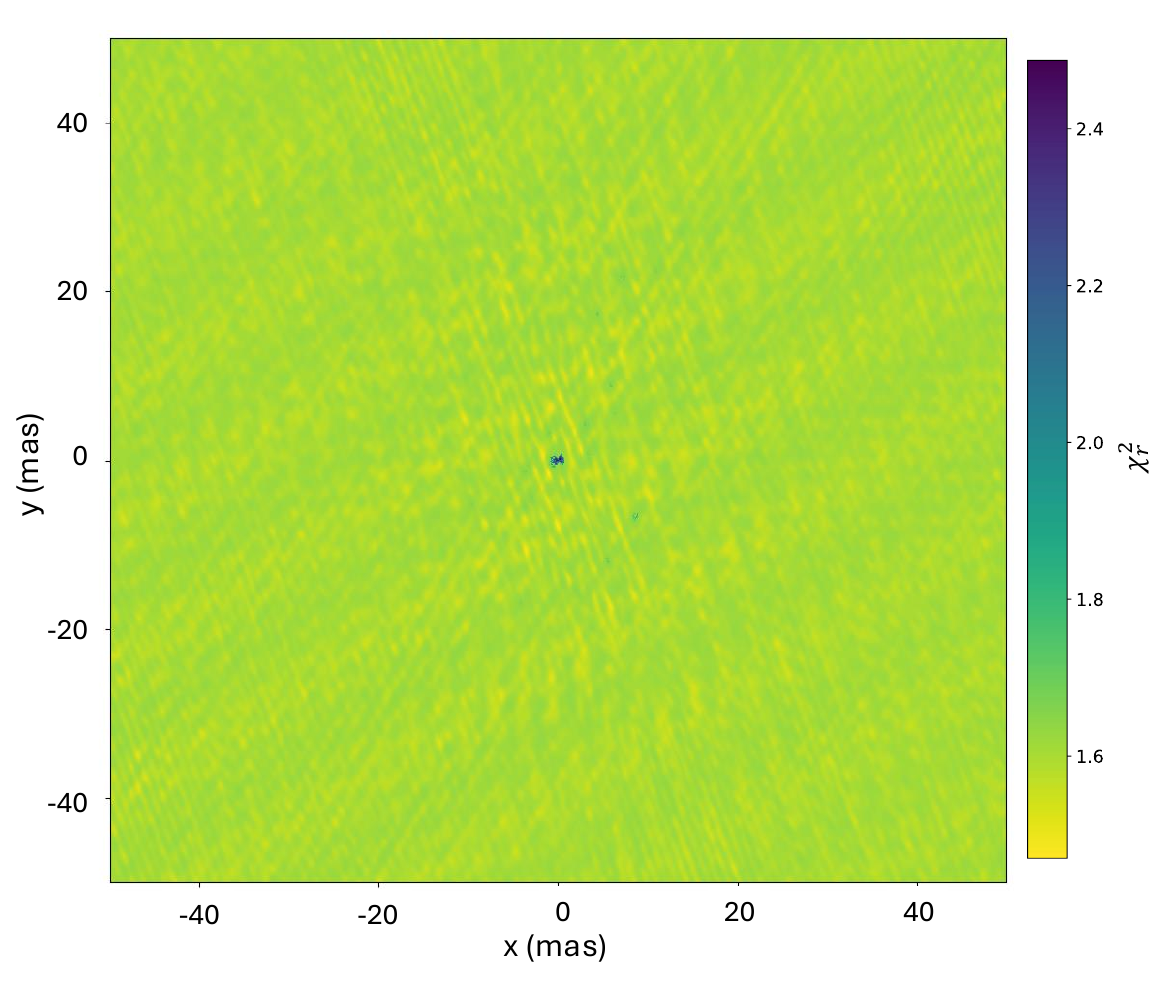}
    \caption{Grid search using the MIRC-X 2021-05-13 dataset using a thin fixed ring model. The grid appears uniform, implying no statistically significant companion detection.}
    \label{fig:grid}
\end{figure}

We initially conducted a search for a stellar companion in a circumbinary disc, as hydrodynamical simulations that included a companion best reproduced the disc seen in scattered light images \citep{Ballabio2021MNRAS.504..888B}. 
A $50$x$50$mas grid search with a step size of $0.1$mas was run on the MIRC-X 2021-05-13 dataset whilst simultaneously using the thin fixed ring model parameters to determine if there was evidence of a companion. The Thin Ring model was used over the Ring R model as we would expect a companion to carve out a cavity and the Ring R model was too wide. We also use the Thin Ring model over the Ring R model as simulations suggest a circumbinary disc. The size of the inner radius for the free ring is much smaller, making the likelihood of a companion existing in this cavity very low \citep{1994ApJ...421..651A,elsender2023MNRAS.523.4353E}. 
This search returns no statistically significant detection within our field-of-view, as seen in Figure \ref{fig:grid}. A caveat to this non-detection is that we are searching in the plane of the disc and it is still possible that a companion is being hidden behind the disc.

% Our grid search (Fig. \ref{fig:grid}) returns a non-detection, ruling out the $0.36 M_\odot$ companion within $0-8$ au proposed by \cite{Ballabio2021MNRAS.504..888B}. 

\begin{figure*}
    \centering
    \includegraphics[width=150mm]{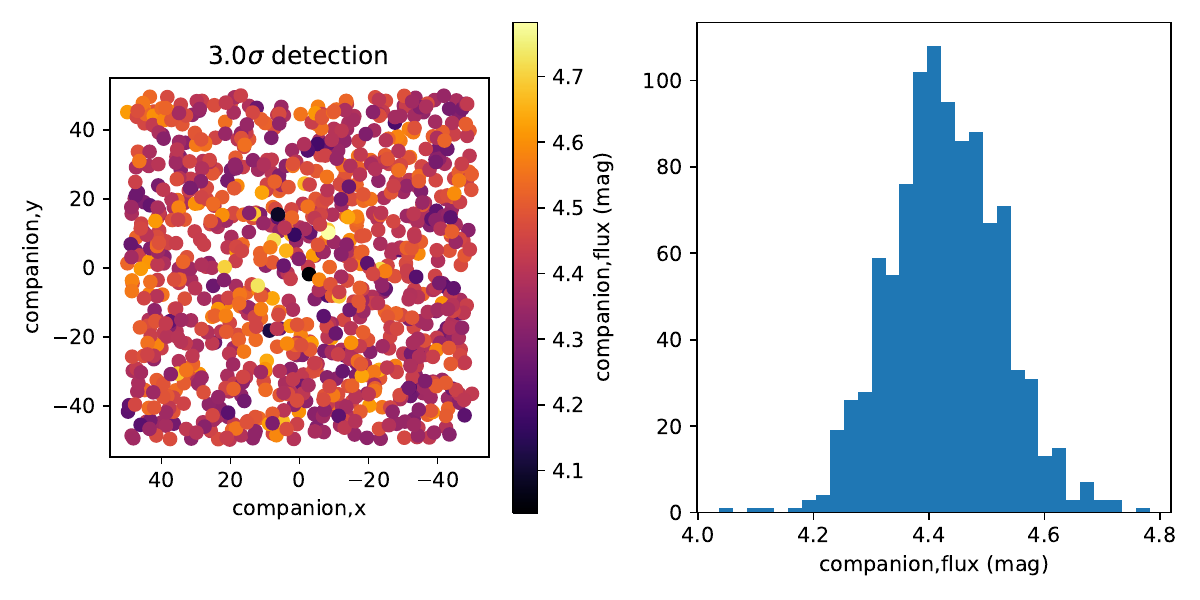}
    \caption{Determining the detection limit of our observations with PMOIRED. \textit{Left: }A random grid where a flux is estimated for a 3$\sigma$ detection. This flux is converted into magnitude for the purpose of this plot. The brighter yellow colours indicate a greater $\Delta$H whilst darker colours indicate a smaller $\Delta$H. The $\Delta$H obtained for our detection limit is $4.43$, corresponding to an upper mass limit of $\sim0.17 M_\odot$. \textit{Right: } Histogram shows the spread of magnitude contrast for our data.}
    \label{fig:injection}
\end{figure*}

As no companion was detected, a detection limit test was performed to place upper limits on the putative companion. Using the chromatic thin ring model (see Figure \ref{fig:injection}) gave us an upper limit in the magnitude contrast of $\Delta$H of $4.5$. PMOIRED implements the method described in \cite{absil2011A&A...535A..68A}, where a random exploration pattern is defined for the companion. At each random point on the grid, a flux is estimated for a 3$\sigma$ detection, then converted to a magnitude. The $\Delta$H contrast corresponds to a stellar mass, which highlights how challenging it is to detect lower mass companions at these separations.

We found the $\Delta$H of $4.5$ corresponded to an apparent magnitude of $12.12$. Using The Cluster Collaboration's Isochrone Server (Version 1.1)\footnote{\url{http://www.astro.ex.ac.uk/people/timn/isochrones/}}, adopting a model interior parameter $\alpha=1.9$ \cite{2002A&A...382..563Baraffe} with the \citet{2011ASPC..448...91Allard} bolometric corrections and additional empirical corrections shown in \citet{2014MNRAS.445.3496Bell},
an upper mass limit for the putative companion was calculated for $5$ Myr, $8$ Myr, and $11$ Myr, using the $\Delta$H found using the detection test and Figure \ref{fig:injection}. This gave us an upper mass limit range of $0.13-0.20 M_\odot$. Further data at longer wavelengths could place tighter constraints on this.

\subsection{Inner disc radius}
\label{sec:innerdiscradius}

From modelling the disc geometry, we are able to confirm the pre-transitional nature of the disc surrounding \hd. The observations are best fit by a broad ring (chromatic) and by an elliptical Gaussian model (achromatic) rather than a narrow ring, suggesting that the inner disc is very broad and emission is not dominated by a narrow inner rim. However, previous studies (eg. \citealt{ibrahim2023ApJ...947...68I}) have shown that a ring model is not complex enough to capture all astrophysical features and we need more data at longer baselines to obtain a more accurate model. In this case, as we lack longer baseline data, we can only fit simple geometries and hence we take the chromatic ring model R as the most physically motivated model (Fig.~\ref{fig:chrom_models}). 

This inner disc radius is $0.26\pm0.04$ mas, corresponding to $0.043\pm0.007$\,au. If we take the inner rim radius from the thin ring model of $0.71\pm0.012$\,mas, this corresponds to $0.117\pm0.002$\,au. \cite{monnier2005} derived a K-band ring radius of $0.81$\,mas, (taking a fractional ring thickness of $20\%$ as adopted in our thin ring model). More recent work by \citet{laz2017A&A...599A..85L} found the inner rim emission to be at a radius of around $0.1$\,au (however the inner rim could not be resolved). Whilst the thin ring model rim radius corresponds well to the emission found by \cite{monnier2005,laz2017A&A...599A..85L} and \cite{benisty2018A&A...619A.171B}, the best-fit ring model gives us a smaller rim radius. \cite{woitke2019PASP..131f4301W} fit an SED for \hd with a stellar excess of around $55\%$, lending more weight to our Ring model being our best fit ($f_{\mathrm{star}}=60\%$), versus our Fixed Thin Ring model ($f_{\mathrm{star}}=70\%$).

We also investigate how the inner rim wall compares with typical sublimation temperatures.
The equilibrium temperature of the inner rim can be calculated using
\begin{align}
\label{eq:subrim}
    R_s & =\frac{1}{2}\sqrt{\epsilon_Q}\Bigg(\frac{T_*}{T_s}\Bigg)^2R_* \\
        & =1.1\sqrt{\epsilon_Q}\Bigg(\frac{L_*}{1000L_\odot}\Bigg)^\frac{1}{2}\Bigg(\frac{T_s}{1500}\Bigg)^{-2} \mbox{ au},
\end{align}
where $\epsilon_Q=Q_\mathrm{abs}(T_*)/Q_\mathrm{abs}(T_s)$ is the ratio of dust absorption efficiencies $Q(T)$ for radiation at colour temperature T of the incident and re-emitted field respectively, and rearranging for $T_s$ (sublimation temperature) \citep{monniermillan2002}
\begin{equation}
    T_s = \left( \frac{1}{R_s}1.1\sqrt{\epsilon_Q}\Bigg(\frac{L_*}{1000L_\odot}\Bigg)^\frac{1}{2}\Bigg(\frac{1}{1500}\Bigg)^{-2}K \right)^{\frac{1}{2}}.
\end{equation}

For the inner edge of the chromatic Ring model, taking $\epsilon_Q \simeq 1.5$ for T Tauri stars \citep{monniermillan2002}, this gives a temperature of $\simeq2300$ K and the chromatic Fixed Thin Ring $T_s\simeq1300$ K. Given that the chromatic Ring model is the best-fit model, this temperature implies that the inner edge is likely to be at the dust sublimation rim. Current estimates on sublimation temperatures range between $1500$\,K to $2000$\,K (eg. \citealt{monniermillan2002,laz2017A&A...599A..85L,gravity2019A&A...632A..53G}), meaning this Ring model has an inner edge temperature that is warmer than expected for $\epsilon_Q$. Similar to the undersized Herbig star conundrum \citep{kraus2015Ap&SS.357...97K}, it may be that there is an ``optically thick" inner gaseous cavity (eg. \citealt{2008ApJ...676..490K,claire2020ApJ...897...31D}). These factors point to the disc being in the pre-transitional stage.

\section{Conclusions}
\label{sec:conclusions}

% Structure the paragraph with something astrophysically exciting at the beginning, e.g.:  You resolve the inner disc that is responsible for the shadows and find the strong mutual misalignment angle.  Then that you distinguish between the 4 cases and determine what sides are facing towards/away from us. Finally that you constrain the scale height of the inner and outer disc. In case there are residuals, this would be an obvious thing to point out for follow-up studies (e.g. to determine whether this point towards real asymmetries or whether a more sophisticated model is needed).

This paper investigates the inner disc geometry surrounding the T\,Tauri star \hd. 
We have resolved the inner disc and determined an inclination angle and position angle of $i=22^\circ\pm 3^\circ$ and $\mathrm{PA}=158^\circ\pm 8^\circ$, with these values being consistent within 1-2$\sigma$ when fitting different geometric models. We find the inner radius to be $0.043$ au.

Using these measurements, we concluded that the inner disc is misaligned relative to the outer disc and thus responsible for casting the shadows seen in scattered light images.  We distinguish between 4 cases of disc alignment and determine that the far side of the inner disc is pointing eastward, and the outer disc has the opposite orientation, with a position angle of the line of shadow of $-15^\circ\pm5^\circ$. From the predicted line of shadow, we can constrain the $h/R$ to be about $13\%$ of the outer cavity wall. This implies a mutual misalignment angle of $\Delta\theta_2=39^\circ\pm5^\circ$ between the inner and outer disc, potentially caused by a large planet on a small inclined orbit, or possibly a flyby event where a warp is propagating through the disc. Observations in the outer disc region may help us to confirm whether a large planet exists in this disc.

We find no evidence for a potential binary companion, but we determine an upper mass limit of $0.13-0.20 M_\odot$ from the $\Delta$H contrast. This contrast corresponds to a stellar mass, which highlights how challenging it is to detect lower mass companions at these separations.  
% Future observations with the CHARA Array including the MYSTIC instrument in the K-band will allow us to constrain the width of the disc.\\
Looking forward, more observations with all six telescopes at CHARA could allow for more data at longer baselines with a more complete uv-coverage. This would enable us to fit a more complex model and do a deeper search for a smaller companion and place tighter constraints on the mass limit derived in this work. Future observations on this object should aim to cover multiple wavelengths to fully constrain the disc and identify the cause of the misalignment. This study could also benefit from following up with radiative transfer modelling to investigate the sublimation rim radius and temperature.

% Looking forward, more observations with all 6 CHARA telescopes could help to get a clearer picture on the morphology of the inner disc, with models chosen that represent a physical disc morphology. 

% something like simulations can only tell us so much, but we need to find the actual evidence i dunno

\section*{Acknowledgements}

We thank the referee for their helpful and useful comments.
We acknowledge support from the European Research Council (ERC) Starting Grant "ImagePlanetFormDiscs" (Grant Agreement No.\ 639889) and Consolidated Grant "GAIA-BIFROST" (Grant Agreement No.\ 101003096). S.K. also acknowledges support from STFC Consolidated Grant (ST/V000721/1).  Travel support was provided by STFC PATT grant ST/S005293/1. MIRC-X received funding under the European Union's Horizon 2020 research and innovation programme (Grant No. 639889) . JDM acknowledges funding for the development of MIRC-X (NASA-XRP NNX16AD43G, NSF-AST 1909165). SM is supported by a Royal Society University Research Fellowship (URF-R1-221669).

Based on observations made with the CHARA Array and with ESO telescopes at the Paranal Observatory (ESO programme IDs 0103.C-0915(A) and 105.20T6.001).

This work is based upon observations obtained with the Georgia State University Center for High Angular Resolution Astronomy Array at Mount Wilson Observatory.  The CHARA Array is supported by the National Science Foundation under Grant No. AST-2034336 and AST-2407956. Institutional support has been provided from the GSU College of Arts and Sciences and the GSU Office of the Vice President for Research and Economic Development.

This research has made use of the Jean-Marie Mariotti Center OiDB service available at \url{http://oidb.jmmc.fr}, as well as the \texttt{PIONIER data reduction package} of the Jean-Marie Mariotti Center\footnote{Available at \url{http://www.jmmc.fr/pionier}}. 

%%%%%%%%%%%%%%%%%%%%%%%%%%%%%%%%%%%%%%%%%%%%%%%%%%
\section*{Data Availability}

VLTI/PIONIER data for programmes 0103.C-0915(A) and 105.20T6.002 are available in the ESO archive. Data for programme 190.C-0963 can be found in OiDB (\url{http://oidb.jmmc.fr/index.html}), using the target name HD 143006. The processed data files will be uploaded to OiDB and the raw and reduced data will be published on the University of Exeter Repository (link will be made available at publication).

%%%%%%%%%%%%%%%%%%%% REFERENCES %%%%%%%%%%%%%%%%%%

% The best way to enter references is to use BibTeX:

\bibliographystyle{mnras}
\bibliography{references} % if your bibtex file is called example.bib

\bsp	% typesetting comment
\label{lastpage}
\end{document}